# Beyond the Dicke Limit: Superradiant J-Exciton-Polaritons in BIC Metasurfaces

Marco Marangi,[1,2] Alexander M. Dubrovkin,[1,2] Anton N. Vetlugin,[1,2] Giorgio Adamo,[1,2*] Cesare Soci[1,2*]

[1] *Centre for Disruptive Photonic Technologies, The Photonics Institute, Nanyang Technological University, Singapore 637371*
[2] *Division of Physics and Applied Physics, School of Physical and Mathematical Sciences, Nanyang Technological University, Singapore 637371*
**Correspondence to: g.adamo@ntu.edu.sg; csoci@ntu.edu.sg*

**Highly correlated photon sources can be realized through cooperative coupling among quantum systems, giving rise to superradiant collective emission. In solid-state ensembles, however, such collective behaviour is typically confined to subwavelength dimensions (the Dicke limit) and strongly suppressed at room temperature by inhomogeneous broadening and rapid dephasing, hindering practical implementations. Here, we show that molecular J-aggregates sustain room temperature superradiant emission and enter a highly collective regime when strongly coupled to the delocalized photonic modes of a silicon bound-state-in-the-continuum (BIC) metasurface. J-exciton polaritons exhibit markedly enhanced, excitation-density-dependent cooperativity, evidenced by an increase in the zero-delay second-order correlation to $g^{(2)}(0) = 3.77$, well within the superbunched emission regime ($g^{(2)}(0) > 2$), and by the growth and reorganization of spatially synchronized emissive domains over areas as large as $8.5\ \mu m^2$, far exceeding the characteristic Dicke area ($\lambda^2$) at the emission wavelength of 589 nm. An interacting-emitter model identifies superradiant clustering as the mechanism by which metasurface-mediated interactions extend J-exciton cooperativity, establishing resonant dielectric metasurfaces as a route towards micrometre-scale superradiance and enhanced photon correlations in disordered materials.**

## Main

Emission from an ensemble of emitters can be drastically modified when they interact through a shared electromagnetic field, giving rise to cooperative optical phenomena such as superradiance (SR) and superfluorescence (SF). In these regimes, emitters synchronize their dipoles and radiate collectively, producing bursts of emission with enhanced emission rates, peak intensities that scale nonlinearly with the number of coupled emitters, and super-Poissonian photon statistics, manifested as strong temporal photon bunching[1].

First introduced by Dicke in 1954, superradiance describes the spontaneous synchronization of an ensemble of emitters through their interaction with a common radiation field within a subwavelength volume[2]. The term superfluorescence was later coined to describe emitters populating an incoherent excited state that must spontaneously synchronize the excited dipoles to build coherence, leading to a time delay between excitation and cooperative emission[3]. Although the terms SR and SF were initially used interchangeably[4], it is now generally accepted that the two phenomena differ in their temporal dynamics and emerge in fundamentally different systems[5,6]: an ensemble of N identical units undergoes SR or SF emission depending on the initial dipole alignment and spatial distribution of the emitting units, thus highly ordered and tightly packed units lead to instantaneous SR emission, while larger inter-unit spacing and disorder result in delayed SF emission. Nowadays, there is resurgent interest in cooperative effects in solid-state emitters[7] as a platform to generate strongly correlated photons[1] for applications in ghost imaging[8,9], multi-photon microscopy[10–13], light detection and ranging[14], non-linear optics[15,16], and quantum key distribution (QKD)[17–19].

In solid-state systems, cooperativity in the form of SF or SR can be observed when emitters align on a microscopic scale through processes of self-assembly, such as ligand-engineering in quantum dot (QD) ensembles[20–26] or solvophobic effects in molecular J-aggregates[27–29]. In QD superlattices the degree of cooperativity is limited by the inhomogeneous broadening of the emitting units and their inter-unit spacing. QD superlattices with typical spacing of ∼10 nm were shown to enter the SF regime at cryogenic temperature and high excitation fluence, displaying redshifted, ultrafast, and delayed bursts of photons with super-Poissonian photon statistics after the onset of dipole phasing. In this regime, the photon statistics are inherently time dependent: during the spontaneous build-up of coherence and the subsequent cooperative emission pulse, the system can display markedly different statistical behaviour, with strongly super-Poissonian fluctuations arising only after dipole phasing and Poissonian statistics being recovered after the superfluorescent burst. Nonetheless, their cooperativity remained limited to just a few tens of units[20]. Shortening of the inter-unit spacing by ligand design has further led to superradiant emission of temporally correlated photons without time delay at cryogenic temperature[22].

Conversely, cooperative systems where emitting units are tightly packed, like molecular J-aggregates with typical inter-unit distances of 1-2 nm [30], can enter the superradiant regime even at room temperature. Upon aggregation, these systems exhibit the hallmark red-shifted J-band and accelerated fluorescence decay associated with cooperative emission. These properties have made J-aggregates exceptional materials to investigate strong light-matter interactions and ultrafast energy exchange mechanisms in microcavities and resonant plasmonic systems[31–33]. Yet, despite their capability of room temperature cooperative emission, disorder and rapid dephasing typically restrict exciton delocalization to only a few molecular units[34].

Recent work on bare perovskite films has highlighted the possibility to further enhance cooperativity of SF systems at high temperature by electron-phonon coupling through the formation of large polarons[35,36], leading to fast delayed emission with large intensity fluctuations[37,38]. This is consistent with complementary theoretical predictions of vibrational decoupling of superradiant excitons in optical cavities[39], where strong coupling with a common optical mode was predicted to reduce the effect of microscale disorder and enable long distance dipole-dipole interactions and macroscopic superradiance[40,41]. Most recently, room temperature synchronization of molecular dipoles was demonstrated in plasmonic nanogap arrays by exploiting collective optical mode coupling to extend spatial coherence[42]. Similarly, exciton polaritons are promising platforms for promoting collective dynamics: the hybridization of many excitons with a common photonic mode can mitigate energetic disorder and generate states whose coherence is governed by the cavity mode rather than solely by the properties of individual emitters. When sufficiently strong relative to the disorder, this hybridization can suppress inhomogeneous dephasing and localization[43–45], and sustain extended phase synchronization even in disordered systems[46–48]. These advances point toward the possibility of engineering long-range excitonic coherence in disordered molecular films by embedding them in nanostructured photonic environments.

In this work, we show that coupling superradiant J-aggregates to resonant dielectric metasurfaces supporting bound states in the continuum (BIC) resonances extends J-exciton cooperativity beyond the Dicke limit. Spatially resolved amplitude and phase maps reveal the formation, growth, and reorganization of locally synchronized emissive domains, supporting superradiant clustering, rather than system-wide phase locking, as the mechanism by which collective emission is enhanced. By identifying the number of independent radiating domains in spatial coherence measurements, we quantitatively link the complex spatial emission landscape to its temporal correlations. Together, these results provide direct evidence of cavity mediated superradiant clustering and establish common-mode coupling as a route to extended cooperativity in disordered molecular ensembles.

## Emergence of superradiance and photon correlations in bare J-aggregate films

To establish a baseline for evaluating cooperative effects induced by BIC metasurfaces, we first characterize the intrinsic optical properties of bare TDBC J-aggregate films. This reference is essential to evaluate metasurface-induced enhancement of superradiance in later sections.

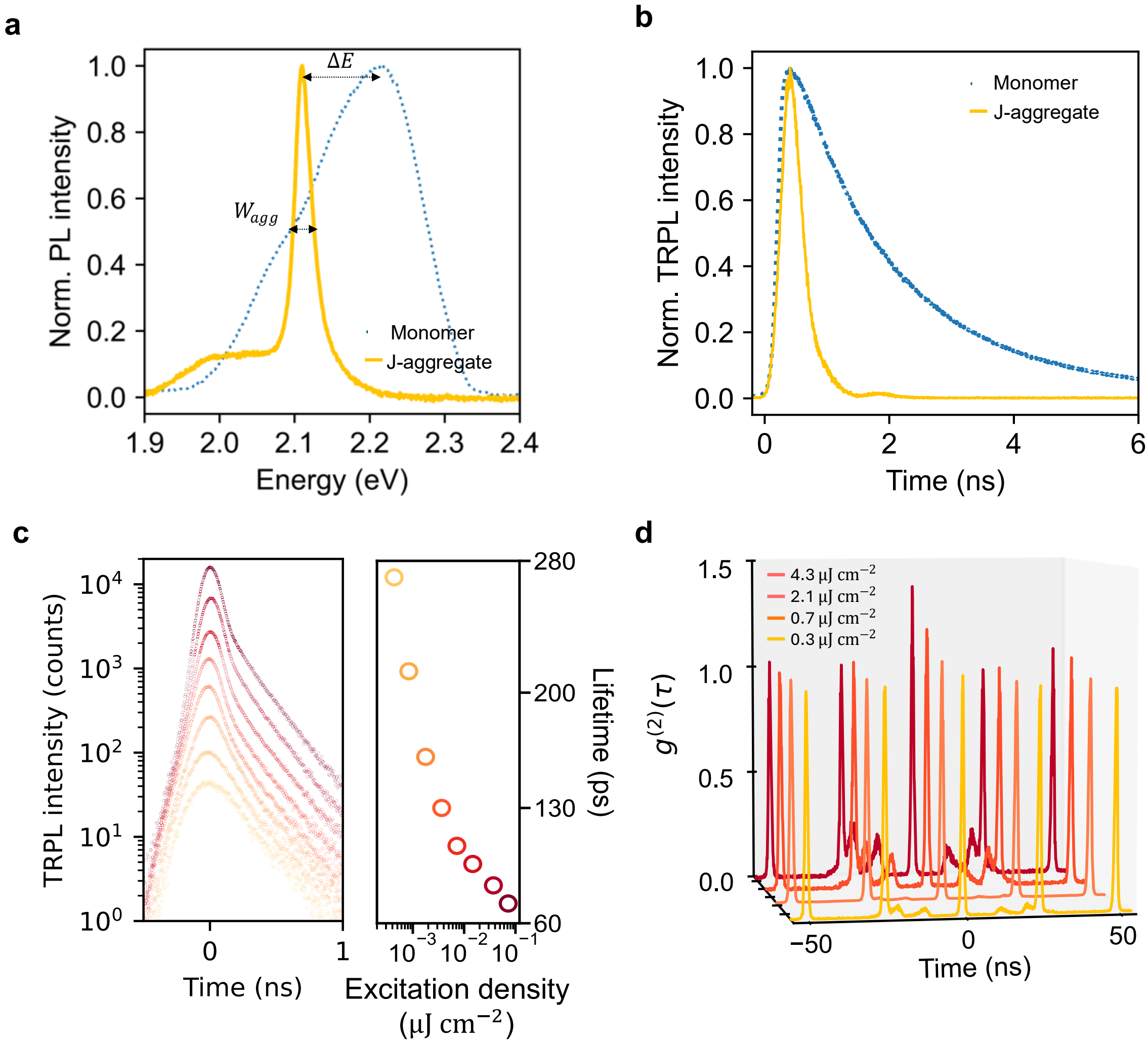

**Fig. 1: J-aggregate emission and photon correlations**. **a**, Photoluminescence spectra of TDBC monomer (blue dotted line) and J-aggregate (yellow solid line) showing the linewidth narrowing and redshift following aggregation. **b**, Time-resolved photoluminescence measurements of monomer and J-aggregate displaying the change in lifetime upon aggregation. **c**, **Left panel:** Photoluminescence time traces of the J-aggregate film as a function of excitation fluence showing the speed-up of the superradiant decay. The time-axis has been normalized to account for experimental time-zero shift of the time controller. **Right panel:** superradiant lifetime of each time trace shortening with increasing excitation fluence. **d**, Second-order correlation function, $g^{(2)}(\tau)$, as a function of excitation fluence for the J-aggregate showing $g^{(2)}(0) > 1$ (photon bunching) at all pump powers.

Solution grown TDBC J-aggregates were deposited on quartz substrates, forming films with a thickness of approximately 25 nm. The photoluminescence (PL) spectrum of the aggregate film differs considerably from that of the monomer film (Fig. 1a): the emission bandwidth narrows from 186 meV for monomers to $W_{agg} = 25$ meV for aggregates, accompanied by a red shift in the peak energy of $\Delta E = E_{mon} - E_{agg} = 104$ meV, with $E_{mon} = 2.21$ eV and $E_{agg} = 2.11$ eV. Both features are characteristic of the formation of J-excitons, arising from strong dipole-dipole coupling between the monomer units, with the emission dominated by the lowest-energy states of the J-band. The coherence length of the J-exciton, $N_{coh}$, can be estimated from the PL spectrum using $N_{coh} = \sqrt{3\pi^2|J|/W_{\mathrm{agg}}} - 1$, where $\mathrm{J} = \Delta E/2.4$ is the dipole-dipole interaction strength[34]. Substituting the measured values, we obtain $N_{\mathrm{coh}} \approx 6$ molecules, consistent with the typical coherence lengths reported for TDBC J-aggregates at room temperature[34].

The formation of J-excitons is further supported by time-resolved PL measurements of the energy filtered exciton peak (FWHM~30 meV). The fluorescence decay is accelerated relative to the monomer emission due to the enlarged oscillator strength of the J-exciton (Fig. 1b). At the lowest excitation fluence of $P_{min} = 4 \times 10^{-4}$ μJ cm$^{-2}$, the J-exciton lifetime ($\tau_{\mathrm{J}} = 270$ ps) reduces to about 5 times that of the monomer film ($\tau_{\mathrm{mon}} = 1480$ ps). At higher excitation fluences the J-exciton time trace can be approximated by two exponential components (Fig 1c): a fast contribution with large amplitude associated with superradiant J-excitons, and a slower weakly emissive component typical of self-trapped, lower coherence excitons (whose contributions are also visible in the low energy shoulder in Fig. 1a) [49]. The fast component of the J-exciton lifetime reduces further to $\tau_{\mathrm{J}} \sim 72$ ps at $P_{max} = 7.22 \times 10^{-2}$ μJ cm$^{-2}$, limited by the temporal resolution of the measuring system. Since the exciton coherence length of J-aggregates is dominated by static structural disorder and remains unchanged under varying excitation conditions, the speed-up of the fast emission component can be attributed to the onset of inter-aggregate superradiance, as predicted for ideal, defect-free, weakly disordered one-dimensional J-aggregates[50]. This implies synchronization of multiple J-excitons, each comprised of $N_{coh}$ coupled units (6 on average), within the common Dicke radiation field.

Temporal photon correlations are a distinctive, but not univocal, feature of superradiant systems, where bright collective modes with large emission rates are formed by coherent coupling of ordered, tightly packed emitters. When few bright collective modes dominate over dark or low emission rate states, the emitted photons follow a super-Poissonian distribution, leading to strong intensity fluctuations and burst-like photon emission[1]. This phenomenon is quantified by the second-order correlation function, $g^{(2)}(\tau) = \frac{\langle \mathrm{I(t)I(t+\tau)} \rangle}{\langle \mathrm{I(t)} \rangle^2}$, where $\mathrm{I(t)}$ is the emitted intensity at time t, $\tau$ is the delay time between two intensity measurements, and $\langle \cdots \rangle$ represents the ensemble average of many emission events. Large intensity fluctuations associated to bursts of photons typical of cooperative emission lead to photon bunching, with $g^{(2)}(0) > 1$. To quantify the temporal correlations in the emission of our J-aggregate films, we measured $g^{(2)}(\tau)$ by computing the all-pair correlation function of all coincidences detected in a Hanbury Brown and Twiss interferometer (see Methods)[51]. At the lowest excitation fluence, the film displays moderate photon bunching at zero-time delay with $g^{(2)}(0) \approx 1.03$ (Fig. 1d). However, $g^{(2)}(0)$ grows superlinearly with increasing excitation fluence, reaching $g^{(2)}(0) \approx 1.32$ when $P = 4.27$ μJ cm$^{-2}$. This observation cannot be explained by conventional (microscopic) J-aggregate SR where cooperativity is assumed to be limited to just a few coherently coupled monomeric units[34]. Rather, the growth of photon correlations at zero delay may stem from an increase of stochastic intensity fluctuations (similar to what was inferred in four-wave mixing experiments in atomic vapours[52], and chaotic quantum dot microlasers[53,54]) in macroscopically coupled J-aggregates. Furthermore, the increase of $g^{(2)}(0)$ in J-aggregate films at high fluence is in stark contrast with the reduction of photon correlations typical of parametric sources[55–57], quantum dot nanolasers, and biexcitonic emission[58,59], where large photon bunching values can be achieved at low fluences, while intensity fluctuations are suppressed in the high excitation regime.

In the following, we show that coupling SR emitters to resonant metasurfaces further enhances inter-aggregate coupling and unlocks a previously unexplored regime of long-range cooperativity at room temperature, resulting in super-thermal temporal correlations ($g^{(2)}(0) > 2$) enabled by micrometer scale spatial coherence.

**J-exciton-polaritons in BIC metasurfaces**

We designed and fabricated three metasurfaces consisting of a square lattice of amorphous silicon ($a$-Si) nanopillars that support a $C_2$ symmetry-protected BIC resonance at the Γ point in the transverse electric (TE) mode. The electric field distribution of the BIC in the x-y and x-z planes is shown in Figs. S2b and S2c, respectively.

The simulated BIC photonic band of the $a$-Si metasurface exhibits the characteristic angular dispersion with quality factor increasing indefinitely (i.e., linewidth narrowing toward zero) toward normal incidence (Fig. 2a), where light cannot couple to the symmetry-protected bound mode (Fig. S2a). This behaviour is consistent with the experimental back focal plane (BFP) TE reflectance map shown in Fig. 2b. The detuning between the photonic mode and the excitonic transition is controlled by varying the nanopillar radius across the different metasurfaces, thereby tuning the spectral position of the BIC band – see reflectance maps of Figs. S3a-c and the corresponding spectral linecuts at θ = 13-17° in Fig. 2c. As expected, the experimental quality factor of $BIC_3$ grows as the reflectance angle approaches θ = 0°, reaching a maximum of ~160 at the smallest accessible angle of incidence of θ~4° (Fig. 2d). In comparison, the quality factor calculated in the presence of material losses is one order of magnitude larger of the measured one. Further reduction of the predicted quality factor can be attributed to structural imperfections due to the limits of nanofabrication[60]. In addition to the BIC mode, the metasurface design also supports a higher-energy Mie-type resonance (Fig. 2), whose interactions with J-aggregates have been investigated in our previous work[61].

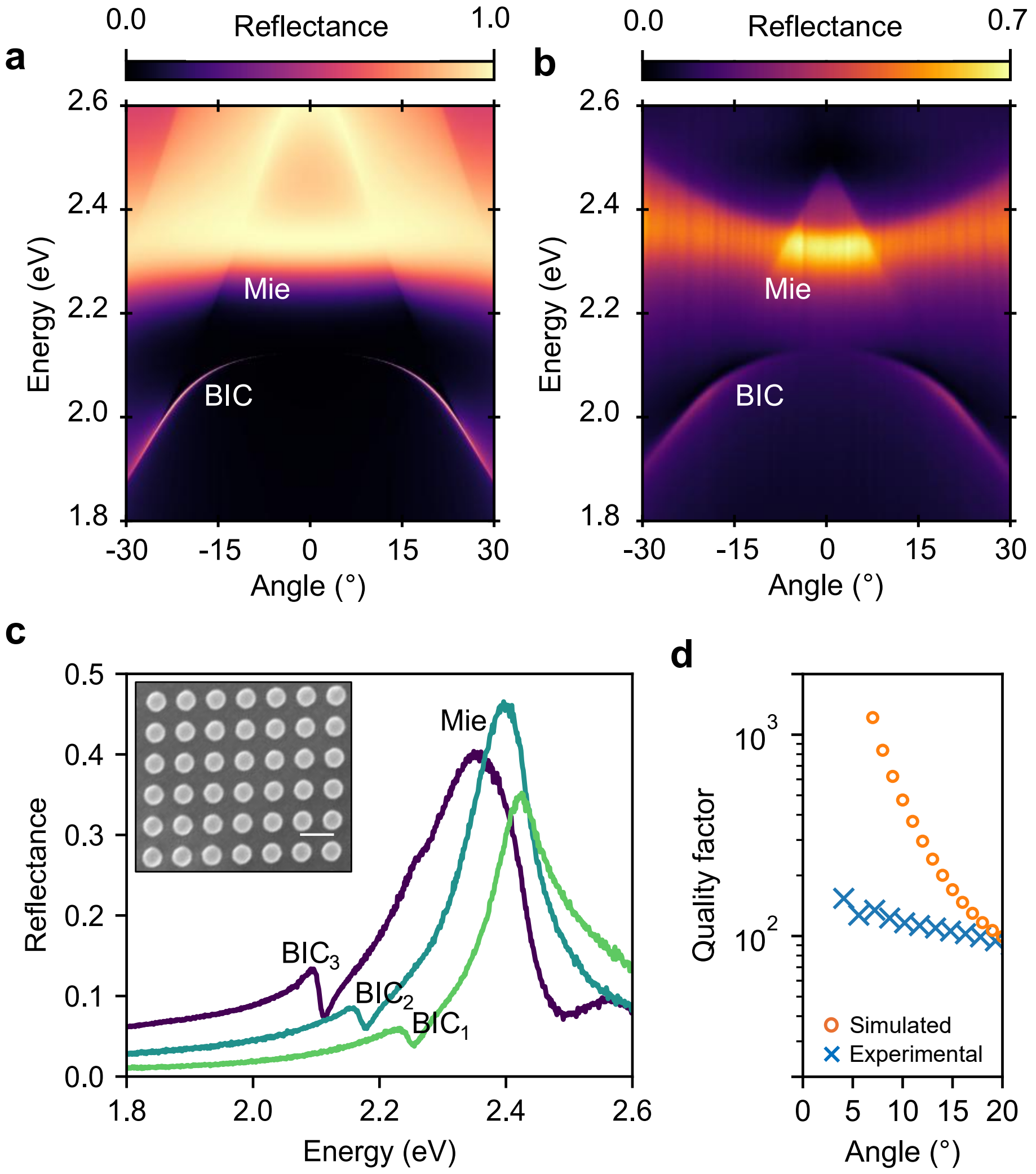

**Fig. 2: Bound-state-in-the-continuum resonance in the a-Si metasurface**. **a**, Rigorous coupled-wave analysis (RCWA) calculated angle-resolved reflectance map showing the high-Q BIC resonance, characterized by a vanishing linewidth toward normal incidence. **b**, Experimental reflectance measured by angle-resolved back focal plane (BFP) Fourier spectroscopy, confirming the presence of the BIC resonance. **c**, Reflectance spectra (linecuts of Figs. S3a-c) for metasurfaces with varying nanopillar radii, illustrating the spectral tunability of the BIC resonance. Inset: SEM image of the Si metasurface; the scale bar is 320 nm. **d**, Quality factor of the $BIC_3$ resonance extracted from the reflectance map in **b** (blue crosses) compared to the simulated quality factor (orange circles).

Each metasurface was coated with a 25-nm thick TDBC J-aggregate film. The geometrical parameters of the metasurfaces, together with the organic film thickness, were optimized to maximize the spectral overlap between the BIC photonic resonance and the J-exciton energy at approximately 2.1 eV. Coating the metasurfaces with the J-aggregate film significantly modifies the photonic band structure, as revealed by the angle-resolved BFP spectral maps for reflectance (Fig. 3a-c) and photoluminescence (Fig. 3d-f). In all cases, a new band emerges below the exciton energy, $J_{exc}$ (indicated by the black dashed line in Figs. 3a,d), signalling the formation of hybrid light-matter states (exciton-polaritons). Fitting the measured dispersion using a coupled oscillator model, with the Rabi splitting energy $\hbar\Omega$ as the sole fitting parameter, confirms that these bands correspond to the lower exciton-polariton branches ($BIC_{1\text{-}3}$ LP, shown as white dashed lines). The corresponding upper polariton branches (BIC UP) are not visible due to absorption losses in both $a$-Si and TDBC at higher energies. The uncoupled photonic band ($BIC_{1,\mathrm{passive}}$) is also indicated for reference by the white dotted line in Fig. 3a. The measured dispersions are in good agreement with full-wave numerical simulations of the coupled system (Fig. S4c).

We extract the energy detuning ($\Delta E$) and Rabi splitting ($\hbar\Omega$) for the three metasurfaces/TDBC systems as follows: $\Delta E_1 = 65.5$ meV with $\hbar\Omega_1 = 305.5 \pm 1.6$ meV (Fig. 3a,d), $\Delta E_2 = 17.7$ meV with $\hbar\Omega_2 = 320.9 \pm 1.6$ meV (Fig. 3b,e), $\Delta E_3 = -28.3$ meV with $\hbar\Omega_3 = 328.0 \pm 1.8$ meV (Fig. 3c,f). The largest Rabi splitting is obtained for the sample with the smallest energy detuning between the photonic mode and the exciton resonance. In all three cases, the condition $\mathrm{g}_0 \gg k_B T \approx 26$ meV is satisfied, indicating that the system is in the strong coupling regime even at room temperature. This is consistent with the requirement from coupled mode theory, $2\mathrm{g}_0 > (\gamma_{\mathrm{exc}} + \gamma_{\mathrm{cav}})/2$, taking the dephasing rate $\gamma_{\mathrm{exc}} = 25$ meV measured from J-aggregate film photoluminescence and the cavity dephasing rate $\gamma_{\mathrm{cav}} = 20$ meV obtained from simulations. At higher energies (~2.0 eV), the system also forms lower polaritons with the Mie resonance at ~2.3 eV (Fig. 2b).

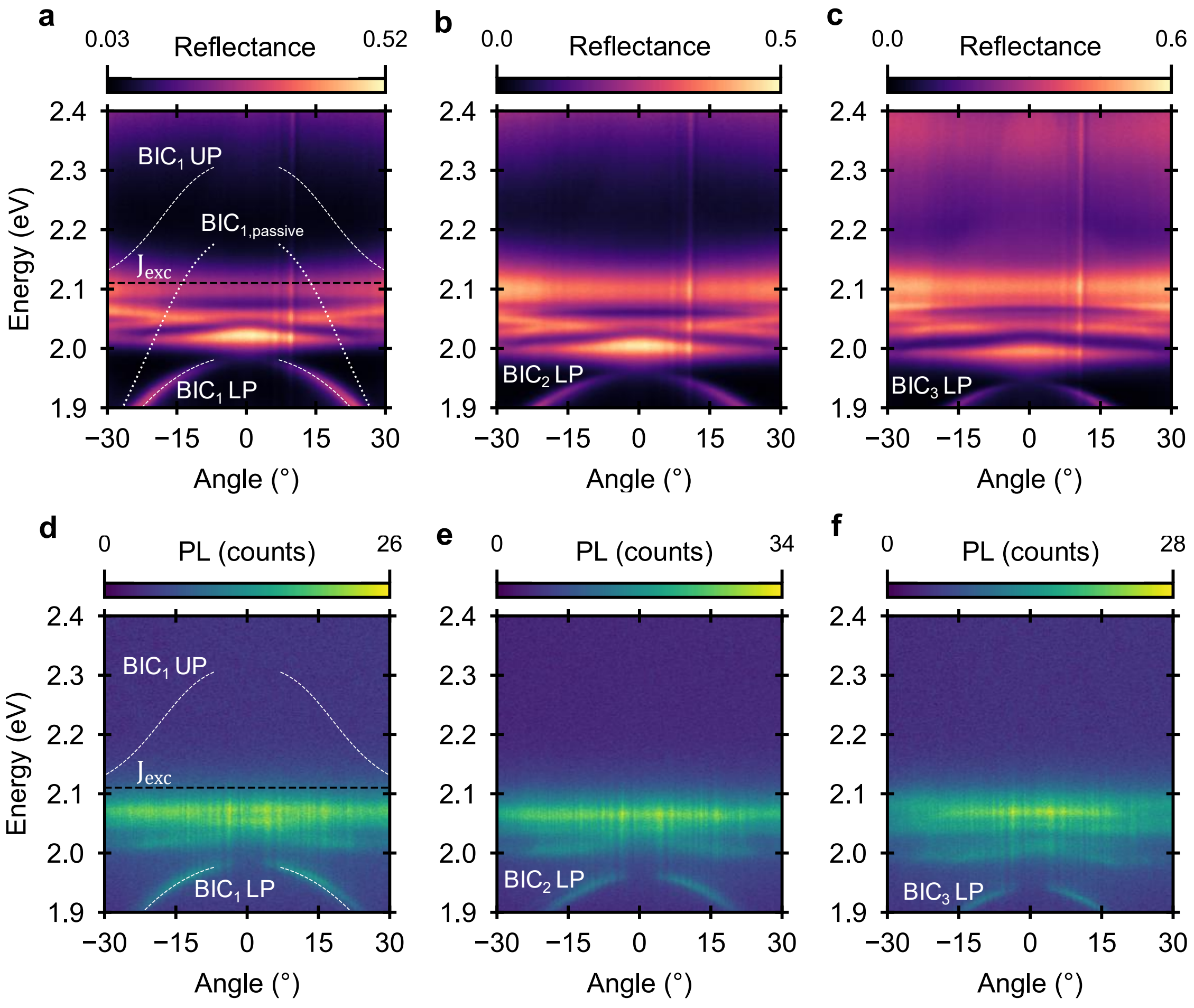

**Fig. 3: Strong coupling of the J-exciton with the bound-state-in-the-continuum (BIC) resonance.** **a–c**, Angle-resolved back focal plane (BFP) reflectance measurements for three metasurfaces with decreasing detuning between the J-exciton ($J_{exc}$, black dashed line) and photonic BIC energies, showing the emergence of a lower polariton (LP) branch below the J-exciton energy. The white dotted lines indicate the uncoupled photonic band ($BIC_1$,passive), while the white dashed lines correspond to the upper and lower polariton branches obtained from a coupled oscillator model fit. **d–f**, Angle-resolved photoluminescence (PL) measurements of the same samples, confirming the formation of the LP branch. The extracted Rabi splitting energies are $\hbar\Omega_{BIC1\ LP}$ = 305.5±1.6 meV, $\hbar\Omega_{BIC2\ LP}$ = 320.9±1.6 meV and $\hbar\Omega_{BIC3\ LP}$= 328.0±1.8 meV, with the largest splitting occurring for the metasurface with the smallest detuning from the exciton energy. The upper polariton branch (UP) is not observed due to absorption losses in Si and TDBC at higher energies.

**Metasurface mediated temporal correlations and large scale cooperativity**

Previous studies have shown that local field enhancement shared across spatially separated J-aggregates can overcome local disorder and promote inter-aggregate exciton delocalization[62]. Here, we leverage the collective and spatially coherent nature of the BIC mode, which extends throughout the entire metasurface, to induce large-scale coherent coupling between distinct J-aggregate domains. To elucidate how resonant metasurfaces promote coupling between spatially separated superradiant emitters and their effect on temporal correlations, we compare the second-order correlation function of photons emitted by J-excitons in bare films and J-exciton-polaritons in coupled metasurfaces (Fig. 4a). The emission of J-exciton-polaritons exhibits larger temporal correlations than uncoupled excitons in the bare film, $g_{LP}^{(2)}(0) > g_{J-exc}^{(2)}(0)$, that systematically increase with coupling strength. In this regime, the correlation function also grows superlinearly when varying the excitation intensity (Fig. 4b). For the largest Rabi splitting and photonic fraction, $g_{LP}^{(2)}(0)$ reaches a maximum value of $3.77$, a threefold enhancement over the uncoupled film at the same pump fluence. This trend parallels the scaling of luminescence with excitation intensity, suggesting that the enhancement of superbunching originates from similar cooperative effects. Note that $g_{SE}^{(2)}(0)$ of the self-trapped excitons, measured at 2.0 eV, remains limited to $g_{SE}^{(2)}(0) < 1.14$ over the range of fluences investigated, ruling out spurious contributions to $g_{LP}^{(2)}(0)$ (see Supplementary Note 5).

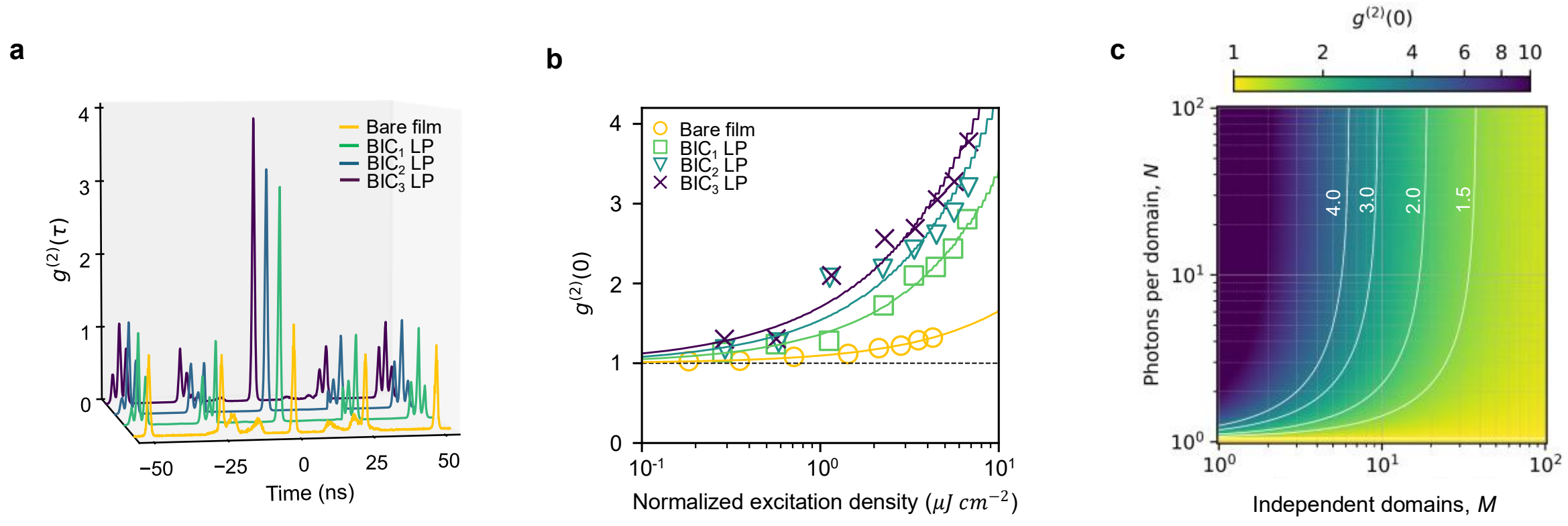

**Fig. 4 J-exciton-polariton superbunched emission. a,** Experimental second-order correlation function, $g^{(2)}_{(\tau)}$, for the $J_{exc}$ and BIC LP emitted photons at the maximum excitation fluence, showing an increase of photon superbunching with growing photonic fraction. The highest bunching value of $g^{(2)}_{(\tau=0)} \sim 3.77$ is obtained for the largest Rabi splitting of $\hbar\Omega_{LP-3} = 328.0 \pm 1.8$ meV. **b,** Measured dependence of the zero delay second-order correlation functions, $g^{(2)}_{(\tau=0)}$, on excitation fluence for the bare film and J-aggregates on BIC metasurfaces with different coupling strengths, together with fitting lines obtained from the model. **c,** Modelled $g^{(2)}_{(\tau=0)}$ for a system consisting of *M* independent domains, each emitting a burst of *N* photons with probability *p=0.02*. A superbunching regime with $g^{(2)}(0) > 2$ is predicted for $M < \frac{N-1}{Np} - 1 < \frac{1}{p} - 1$.

To quantify the enhancement of cooperativity induced by the metasurface, we relate the second-order correlation function to the degree of inter-aggregate synchronization of J-exciton emitters. We assume that the excitation area can be partitioned into *M* independent macroscopic domains, each containing *N* J-excitons. Each exciton acts as an effective single-photon emitter, and excitons in different domains are assumed to be similar in their spectral properties and geometric parameters (e.g., transition dipole orientation). Within each domain, emitters will either synchronize and emit collectively in the form of a burst of *N* photons with probability $\Phi$ (macroscopic superradiance), or remain dark with probability $1 - \Phi$. This picture leads to a binomial photon-number distribution across the *M* domains, from which the zero-delay second-order correlation function is

$$g^{(2)}(0) = \frac{\langle n(n-1)\rangle}{\langle n\rangle^2} = \frac{M-1}{M} + \frac{N-1}{MN\Phi}, \qquad (1)$$

and the number of independent domains *M* can be obtained inverting Eq. 1:

$$M = \frac{N-1-\Phi N}{\Phi N[g^{(2)}(0)-1]}. \qquad (2)$$

*M* provides a quantitative measure of the spatial extent of macroscopic synchronization within the excitation area: for a given pump fluence, the increase of $g^{(2)}(0)$ observed in strongly coupled BICs (growing with Rabi splitting) implies a smaller number of independent domains $M$, thus extended inter-aggregate synchronization over macroscopic distances. Fig. 4c shows the values of $g^{(2)}(0)$ computed with Eq. 1 as a function of $M$ and $N$ for a fixed value of $\Phi = \Phi_{TDBC} \sim 0.02$ [49,63] (the contour lines correspond to experimentally relevant values of $g^{(2)}(0)$). Overall, strong photon superbunching is found when the number of independent domains is small. In this regime, the emission is dominated by rare but intense superradiant bursts, which generate large intensity fluctuations and therefore high $g^{(2)}(0)$. The model also shows that for large $N$, the correlation function becomes insensitive to the number of emitters per domain: for $N \gg 1$, the second term in the expression approaches $1/(\Phi M)$, independent of $N$. These trends indicate that strong cooperativity, and thus highly superbunched photon emission, emerges only when a small number of large, highly synchronized domains contribute to superradiance. Since in the bare film inter-aggregate coupling occurs within subwavelength distances, here taken as $\lambda_{sub} = 0.3\lambda_{J-exc} \approx 180$ nm [1], we calculate the number of absorbed photons per subwavelength area, $N$, and use Eq. 2 to estimate the number of independent domains $M$. We find that $M$ decreases systematically with excitation fluence and photonic fraction (solid lines in Fig. 4b). For fixed $N$, the decrease in $M$ together with the increase in $g^{(2)}(0)$ in the coupled system suggests that the metasurface extends the scale of inter-aggregate cooperativity.

Although superlinear emission and photon bunching are widely regarded as signatures of cooperative emission, neither can be uniquely attributed to collective dynamics, as amplified spontaneous emission, biexcitonic emission, and near-threshold polariton condensation can produce similar signatures[26,64,65]. To elucidate the mechanism underlying the enhanced photon correlations, we therefore measured the first-order spatial coherence of emission from both the bare J-aggregate film and exciton-polaritons (see Methods for details). While the bare film does not show evidence of increased spatial coherence and is limited to subwavelength coherence (Fig. S13a,e), the J-exciton-polaritons emission shows the formation of coherence fringes spanning micrometer length scales (Fig. S13b-d).

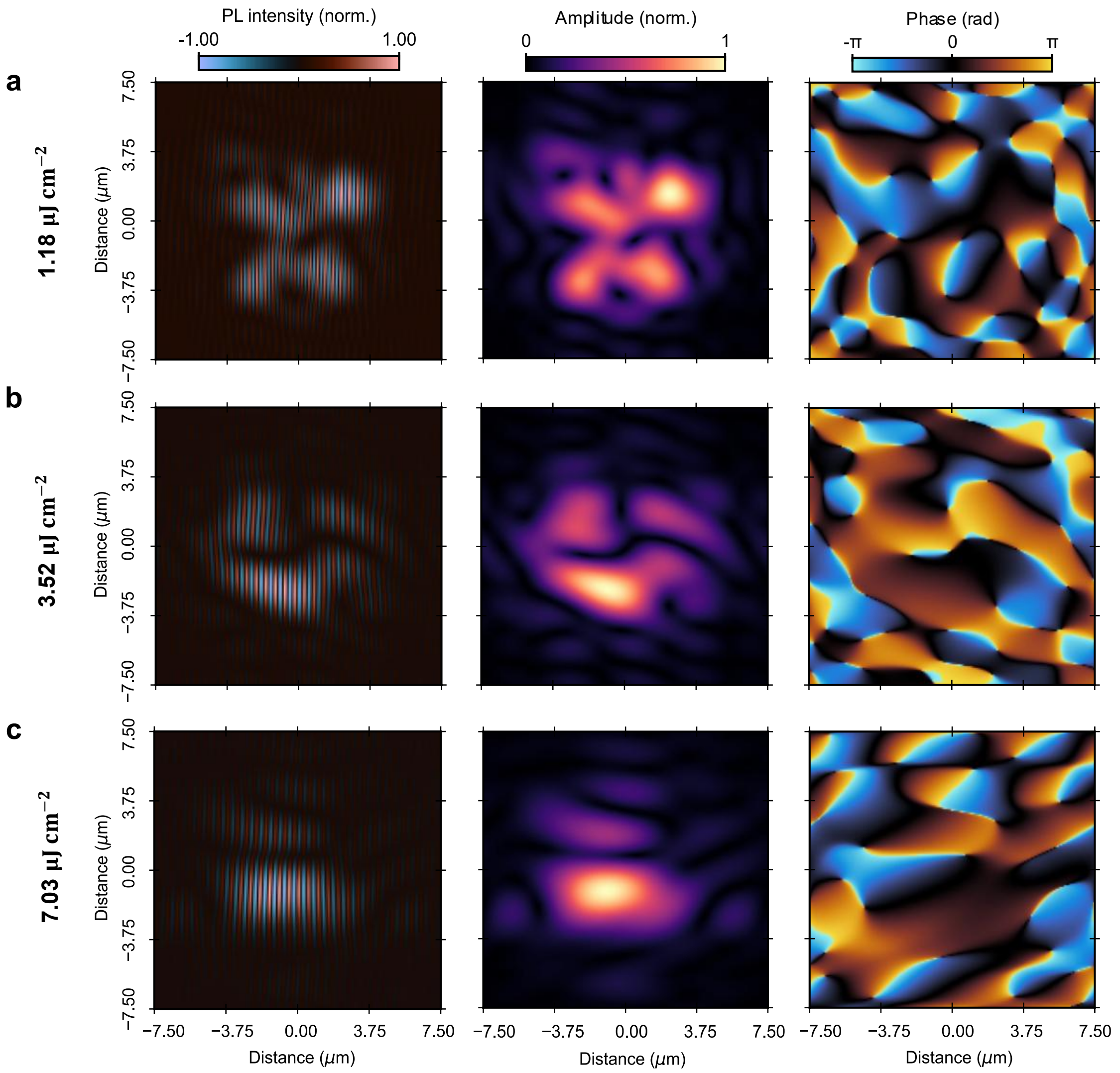

**Fig. 5: Emission topology of superradiant domains. a-c,** The evolution of superradiant clustering with increasing fluence is shown for BIC $LP_3$ via the fringe (left panel), amplitude (middle panel), and wrapped phase (right panel) maps. As the fluence increases the number of coherent domains decreases while their size increases.

Amplitude and phase topology maps of the measured interference patterns (Fig. 5) provide a steady-state visualization of inter-aggregate emission, where spatially extended superradiant domains appear as regions of high amplitude and slowly varying phase. Increasing the excitation density changes the emission landscape of BIC $LP_3$: at the lowest fluence, many small and isolated domains are separated by amplitude and phase discontinuities (Fig. 5a); while at larger fluences (Fig. 5b-c), larger connected domains with more uniform phase become increasingly more prevalent. The evolution of inter-aggregate superradiance with fluence was systematically tracked applying a combined amplitude- and phase-based domain selection algorithm (see details in Supplementary Note 7). This analysis allows a quantitative evaluation of the proposed superradiant clustering mechanism confirming that the number of independent emissive domains decreases significantly with increasing excitation density (blue markers in Fig. S15a), going from 38 to 20 domains on average at the lowest and highest excitation densities measured, respectively. Concurrently, the global phase-order parameter[66], $R_{\phi} = |\langle e^{i\phi}\rangle|$, is maximized at the highest excitation density but does not reach unity (inset of Fig. S15a), indicating metasurface-enhanced but incomplete global phase-locking consistent with the presence of multiple independent emissive domains. The number of domains extracted from spatial correlation measurements are in good agreement with those predicted by the $g^{(2)}$ model (purple crosses in Fig. S15a) at the probed excitation densities, confirming the validity of the model. Furthermore, calculating the domain-area survival distribution (Fig. S15b) shows that the probability to find large domains grows systematically with excitation density, leading to a maximum mean area of $8.5 \pm 1.5$ μm$^2$. The increasingly prominent high-area tail confirms that the decrease in domain number is driven primarily by the formation of a few large clusters, rather than by a uniform increase in the size of all domains, consistent with superradiant clustering.

## Discussion

The excitation density dependent intra-domain synchronization of J-exciton-polaritons (Fig. 4) and the accompanying domain growth (Fig. 5a,b) can be understood in analogy with synchronization transitions in disordered polariton lattices and condensate networks[43,67,47]. In these systems, spatially localized condensates with initially distinct phases and energies develop an extended phase-locked state when intersite coupling overcomes disorder-induced detuning[47,46,68]. Here, each J-aggregate superradiant domain acts as a local collective oscillator, while the delocalized BIC-polariton field mediates coupling between domains beyond the range of their intrinsic subwavelength interactions. Unlike synchronized polariton lattices, here the system does not develop global phase locking but remains partitioned into locally synchronized domains (Fig. 5 and inset of Fig. S15a). This behaviour is consistent with the strong morphological disorder and short temporal coherence of the J-aggregate emission and more closely resembling the local synchronization recently reported in plasmonic nanoarrays[42]. Crucially, these partially synchronized states retain Dicke-like emission, such that growing spatial coherence is accompanied by increasing photon superbunching rather than Poissonian photon statistics.

Together, these results show that strong coupling between J-excitons and a bound-state-in-the-continuum resonance of a dielectric metasurface enhances room-temperature superradiant cooperativity beyond the Dicke limit. Comparing the independently measured effective number of domains with the values derived from a $g^{(2)}$ model of interacting emitters, we have quantitatively linked the evolution of spatial coherence to temporal correlations. This correspondence provides real space evidence of superradiant clustering and demonstrates how metasurface-mediated common-mode coupling can generate strong spatial and temporal correlations in disordered systems. The resulting excitation-dependent correlated emission may find application in correlated imaging[11–13], ranging[14], and thermal-light communication protocols[18,19].

## Methods

### Metasurface design

The metasurface supporting a bound-state-in-the-continuum resonance was designed via Finite Difference Time Domain (FDTD) and Rigorous Coupled-Wave Analysis (RCWA) suite of the commercial software Ansys Lumerical. The nanoresonator geometry supports an out-of-plane magnetic dipole resonance that cannot couple to an incoming transverse electric excitation at normal incidence, giving rise to a symmetry-protected BIC at $\theta = 0°$. Comsol Multiphysics eigenfrequency solver is instead utilized to calculate the quality factor of the BIC as a function of incident angle (Fig. S2a).

### Metasurface fabrication

Quartz substrates were cleaned by sequential sonication in solvents: AR 600-71 (Allresist), acetone and IPA. The substrates were then dried with nitrogen and underwent a short oxygen plasma treatment. Undoped amorphous Si layer was grown using a plasma enhanced chemical vapor deposition system (Cello Aegis-20) at 200°C. 50 µm x 50 µm BIC metamaterial arrays were defined using e-beam lithography with AR-N 7520 resist and Electra 92 conductive coating (Allresist), followed by inductively coupled plasma reactive ion etching of Si with SF6/O2 (Oxford PlasmaPro100 ICP-RIE system). The remaining resist was removed by sequentially soaking the sample in solvents: AR 300-76 (Allresist), AR 600-71, acetone and IPA. The samples were then dried with nitrogen, followed by a short oxygen plasma treatment.

### Processing and deposition of TDBC J-aggregates

The cyanine dye 5,6-Dichloro-2-[[5,6-dichloro-1-ethyl-3-(4-sulfobutyl)-benzimidazol-2-ylidene]-propenyl]-1-ethyl-3-(4-sulfobutyl)-benzimidazolium hydroxide, inner salt, sodium salt J-aggregate (TDBC) was purchased from Biosynth Carbosynth (CAS: 18462-64-1) and was dissolved in de-ionized water with a concentration of $20\ \frac{\mathrm{mg}}{\mathrm{mL}}$. The solution was left in the dark for 3 days to form J-aggregates. 100 μL of the solution were then deposited by spincoating at 4000 rpm for 60 seconds on the fabricated Si metasurfaces. The sample was then dried with $N_2$.

**Angle-resolved spectroscopy**

Angle-resolved reflectance (ARR) and photoluminescence (ARPL) measurements were performed with a custom-built microspectrometer setup consisting of an inverted optical microscope (Nikon Ti-u, 50x objective, NA = 0.55), a spectrograph (Andor SR-303i with a 300 lines/mm grating), and a charge-coupled detector (CCD, Andor iDus 420). A series of lenses along the beam path between the microscope and the spectrograph projects the back focal plane (BFP) of the collection objective on the slit of the spectrograph, allowing the collection of angular information within bounds defined by $\frac{k_x}{k_0} = \mathrm{NA} = 0.55$. The sample was excited via a halogen lamp for ARR measurements, while a 405 nm continuous wave laser was used for ARPL measurements.

**Photoluminescence and photon correlation spectroscopy**

Steady-state, time-resolved, and photon correlation measurements are performed by exciting the sample with a 405 nm picosecond pulsed laser with 40 MHz repetition rate (PicoQuant PDL 800-D). The emission of the sample is collected via a 100x magnification 0.9 NA microscope objective and is spectrally filtered by interchangeable bandpass filters. A dichroic mirror and a longpass filter are utilized to remove the reflected and scattered laser beam. After filtering, the emission can be redirected to a spectrometer or a laser-synched time controller (PicoHarp300) via fibers, to perform steady-state or time-resolved and photon correlation measurements, respectively. An avalanche photodetector with <50 ps timing resolution (MPD PDM) is utilized for time-resolved photoluminescence measurements.

The photon statistics discussed in the manuscript are measured in a standard Hanbury-Brown-Twiss (HBT) interferometer. The emission of the sample is collected via a microscope objective and is re-directed to a 50/50 fiber beam splitter connected to two avalanche photodetectors (Excelitas SPCM-AQRH-14-FC, APD 1 and APD 2, with <350 ps single photon timing) for single-photon counting. A time controller (ID Quantique ID900) records the raw timestamped photon arrival times collected by the detectors and builds the second-order correlation function, $g^{(2)}(\tau)$, as calculated by the all-pair correlation function[51]. During all measurements, the photon count rate has been kept below 5% of the laser repetition rate to prevent pile-up. The resulting $g^{(2)}(0)$ is normalized by dividing by the average peak area of all side peaks, where emission is uncorrelated.

**Michelson interferometry**

The spatial coherence of emission is measured via a Michelson interferometer (Fig. M1) where the J-exciton-polariton emission is pumped by a 515 nm picosecond pulsed diode laser (QuixX 515-80 ps), collected by a 50x (0.55 NA) objective, filtered via a 10 nm bandpass filter, and split into two arms by a 50/50 beamsplitter. The emission in the two arms is time-matched via a motorized stage before being reflected back into the beamsplitter by a mirror (arm 1) and a retroreflector (arm 2). The emission is then recombined on a CCD (Andor iDus 420C) to form interference fringes in the real space image. The fringes profiles are extracted from the interferogram by Fourier filtering the high frequency components. The spatial coherence is then calculated by fitting the fringe envelope via a Gaussian and extracting its full width at half maximum (FWHM).

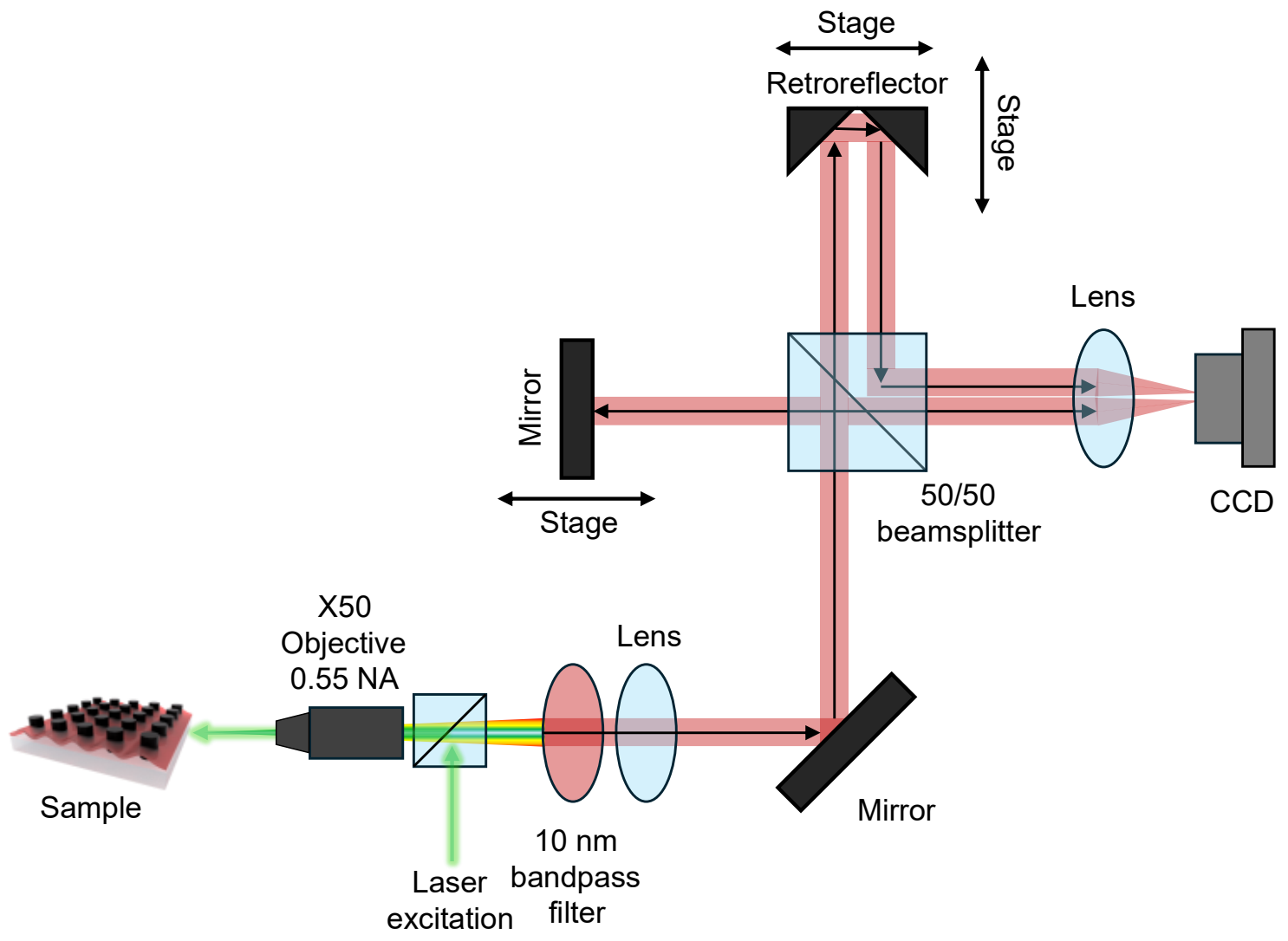

**Figure M1:** Schematic of the Michelson interferometer used to measure the spatial coherence of the J-exciton-polariton emission.

## Author contributions

M.M., G.A., and C.S. conceived the idea. M.M. performed all simulations, spectroscopy measurements, and analysed the data with the help of G.A.. A.M.D. carried out design optimisation and fabrication of the silicon metasurfaces. A.N.V. modelled the superradiant emission from interacting J-aggregates and their second-order correlation function. M.M. analysed the amplitude and phase maps extracted from the spatial correlation measurements. M.M., G.A. and C.S. drafted the manuscript, with all authors contributing to editing and review. C.S. supervised the work.

## Acknowledgements

Research was supported by the Singapore Ministry of Education, (Grant no. MOE-T2EP50222-0015) and by the Singapore's National Research Foundation through the National Centre for Advanced Integrated Photonics (Grant no. NRF-MSG-2023-0002). The authors would like to acknowledge and thank the Nanyang NanoFabrication Centre ($N^2FC$).

**Supplementary information**

# Beyond the Dicke Limit: Superradiant J-Exciton-Polaritons in BIC Metasurfaces

Marco Marangi,[1,2] Alexander M. Dubrovkin,[1,2] Anton N. Vetlugin,[1,2] Giorgio Adamo,[1,2*] Cesare Soci[1,2*]

*[1] Centre for Disruptive Photonic Technologies, The Photonics Institute, Nanyang Technological University, Singapore 637371*
*[2] Division of Physics and Applied Physics, School of Physical and Mathematical Sciences, Nanyang Technological University, Singapore 637371*
**Correspondence to: g.adamo@ntu.edu.sg; csoci@ntu.edu.sg*

## Table of Contents

**Supplementary Note 1: J-aggregate film characterization**

Roughness and thickness of J-aggregate films are characterized by atomic force microscopy (AFM). Films spin coated on glass substrates have absolute roughness of 3.15 nm and thickness of 25 nm after (**Figs. S1a-c**). The complex refractive index of the film determined by ellipsometry (**Fig. S1d**) is used in the RCWA simulations to calculate the photonic band dispersion of the *active* structure.

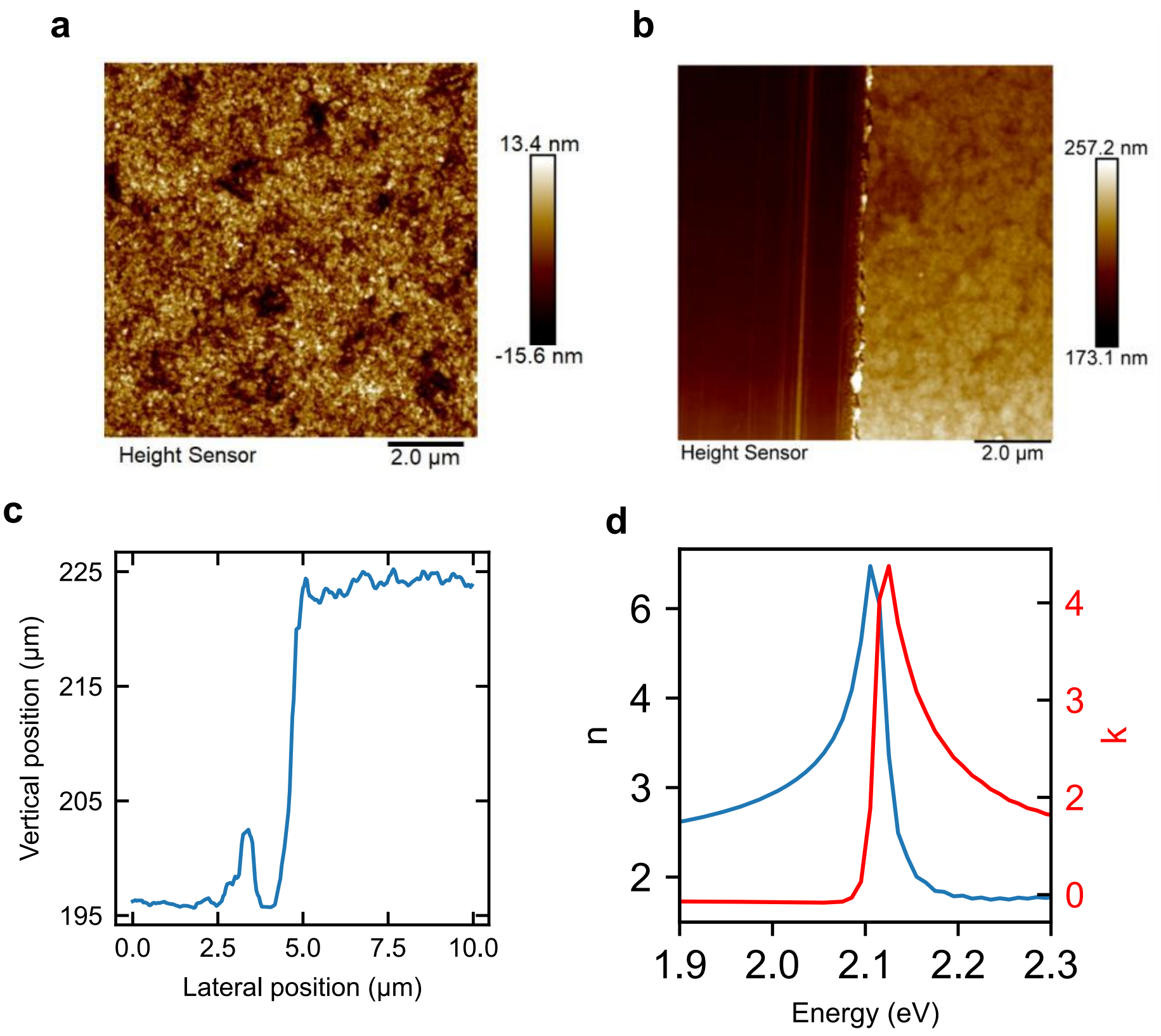

**Fig. S1: Bare J-aggregate film characterization. a**, AFM surface scan of a J-aggregate film with absolute roughness of 3.15 nm. **b**, AFM surface scan of a J-aggregate film around a cut made with a syringe tip to characterize the film thickness. **c**, Profile measurement along a horizontal cut of **b**, revealing a J-aggregate film thickness of 25 nm. **d**, Measured refractive index of the J-aggregate film.

## Supplementary Note 2: Bound-state-in-the-continuum resonance in silicon metasurfaces

The silicon metasurfaces supports a 2-fold rotational ($C_2$) symmetry-protected BIC, arising from an out-of-plane magnetic dipole (MD), generated by displacement currents around the normal direction. Away from the normal direction, the bound states become leaky with finite quality factors as reported in **Fig. S2a**. The electric field distribution of the mode is shown in **Figs. S2b and S2c**, depicting large field enhancements in proximity to the TDBC film and a null electric field at normal incidence as expected of BICs.

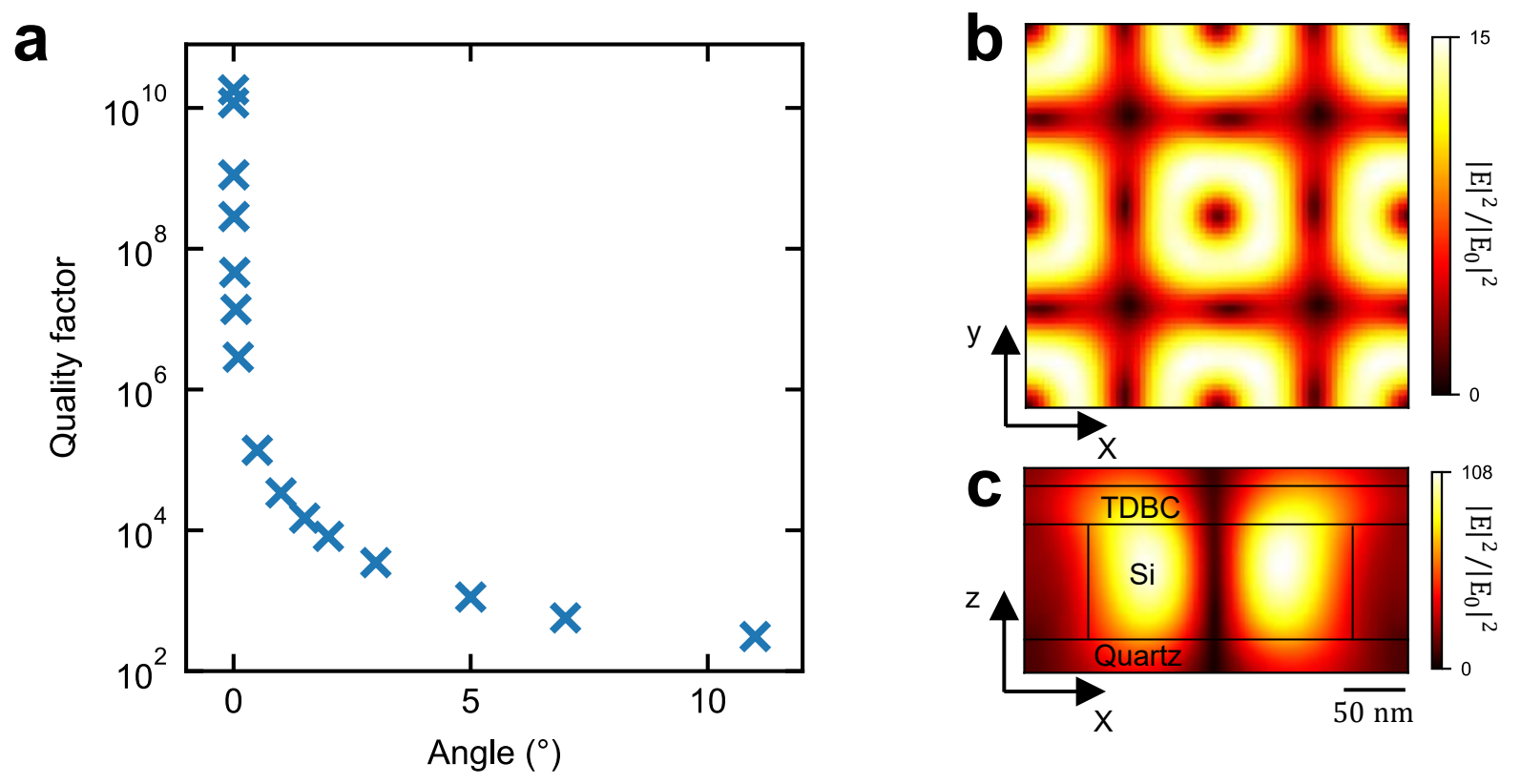

**Fig. S2: Quality factor and field enhancement of the bound state in the continuum (BIC) resonance. a,** Quality factor of the BIC as a function of angle, shown to increase exponentially as the resonance approaches the normal direction (0°). **b-c**, Normalized electric field magnitude spatial distribution in a horizontal (**b**) and (**c**) vertical cross section showing large field enhancement in proximity to the TDBC film. The horizontal cross section is taken in the middle of the TDBC film in the x-y plane, while the vertical cross section is taken at the centre of the unit cell cutting the Si nanopillar in half along the x-z plane.

The experimental photonic dispersion of the metasurfaces discussed in the main text are shown in **Figs. S3a-c**, where the BIC appears as a Fano-shaped peak in reflectance with vanishing linewidth as the dispersion approaches the symmetry-protected normal incidence. The BIC resonance is redshifted by increasing the radius of the nanopillars between 72 to 78 nm (as reproduced by RCWA simulations shown in **Fig. S4a** and Fig. 2a), while the thickness is set to 100 nm and the unit cell period is kept constant at 315 nm. The BIC appears more defined as the radius increases due to its spectral position approaching a spectral range where silicon has lower losses, in accordance to the results reported in [1].

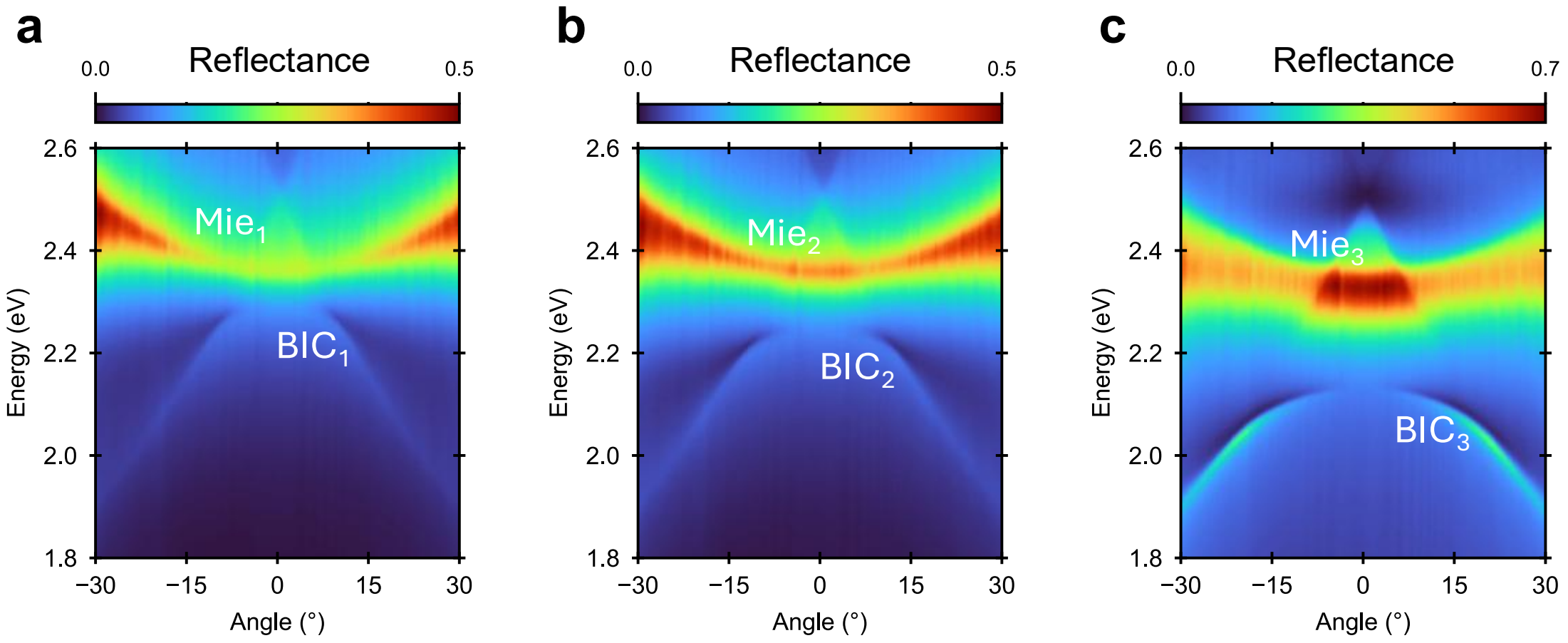

**Fig. S3: Experimental photonic dispersion of the bare BIC. a-c,** Experimental angle-resolved reflectance of the bare metasurfaces showing the typical BIC dispersion for increasing radii of the nanopillars.

## Supplementary Note 3: Strong coupling regime

To confirm that the *active* system is in the strong coupling regime, a coupled oscillator model is used to model the J-exciton and the cavity resonance as two interacting harmonic oscillators that give rise to the upper and lower exciton-polariton branches. The J-exciton is modelled as an oscillator with fixed, dispersion-less, resonance energy $E_{exc} = 2.11\ eV$ and losses defined by its linewidth $\gamma_{exc}$=0.025 eV. While the BIC cavity resonance energy $E_{cav}$ and losses $\gamma_{cav}$ are both highly dispersive as shown in the photonic dispersion in Fig. 2a. The photonic dispersion of the BIC interacting with the J-exciton in the *active* system differs from that of the bare BIC due to the presence of the J-aggregate thin film due to its refractive index (see Fig. S1d). To account for this, the photonic dispersion utilized in the coupled oscillator model corresponds to what is generally referred to the photonic dispersion of the *passive* system, where the bare metasurface is simulated while covered by a 25 nm thick J-aggregate film with a fixed refractive index n=2.2 (taken from the experimental refractive index of the J-aggregate). As shown in **Figs. S4a and S4b**, the bare and passive photonic dispersion ($BIC_1$) are similar in shape but differ in energy as the refractive index of the J-aggregate film causes a red-shift of the whole dispersion. While the photonic band structure of the active metasurface (Figs. S4c and S4d for the calculated and experimental dispersion, respectively) is redshifted in energy and distorted in dispersion, hinting at strong light-matter interactions between the J-exciton and $BIC_1$, and the observation of the lower polariton band ($BIC_1$ LP).

To confirm this hypothesis, the interactions between the J-exciton and the BIC can be modelled by a coupled oscillator model, which predicts the formation of the upper and lower exciton-polariton bands:

$$E_{LP,UP} = \frac{1}{2}\left[E_{exc} + E_{cav} + \frac{i(\gamma_{cav} + \gamma_{exc})}{2}\right] \pm \sqrt{g^2 + \frac{1}{4}[E_{exc} - E_{cav} + i(\gamma_{cav} - \gamma_{exc})]^2}$$

Plugging the energy and losses of the J-aggregate, and the simulated passive cavity dispersion into the coupled oscillator model leaves the coupling strength, *g*, as the sole fitting parameter. The coupling strength is thus determined by minimizing the error between the calculated lower polariton dispersion and the experimental lower polariton emission. The Rabi splitting, $\hbar\Omega$, is calculated from the coupling strength based on the formula $\hbar\Omega = 2\sqrt{g^2 + \frac{(\gamma_{cav} - \gamma_{exc})^2}{4}}$.

The calculated upper and lower polariton bands are showed Fig. S4d, with the lower polariton matching the experimental dispersion when the Rabi splitting is calculated to be $\hbar\Omega_{LP\text{-}1}$ = 305.5±1.6 meV, while the upper polariton is not observed due to the strong absorption of the TDBC film in the predicted spectral region. The same fitting procedure is performed for the two other metasurfaces, yielding Rabi splittings of $\hbar\Omega_{LP\text{-}2}$ = 320.9±1.6 meV and $\hbar\Omega_{LP\text{-}3}$ = 328.0±1.8 meV.

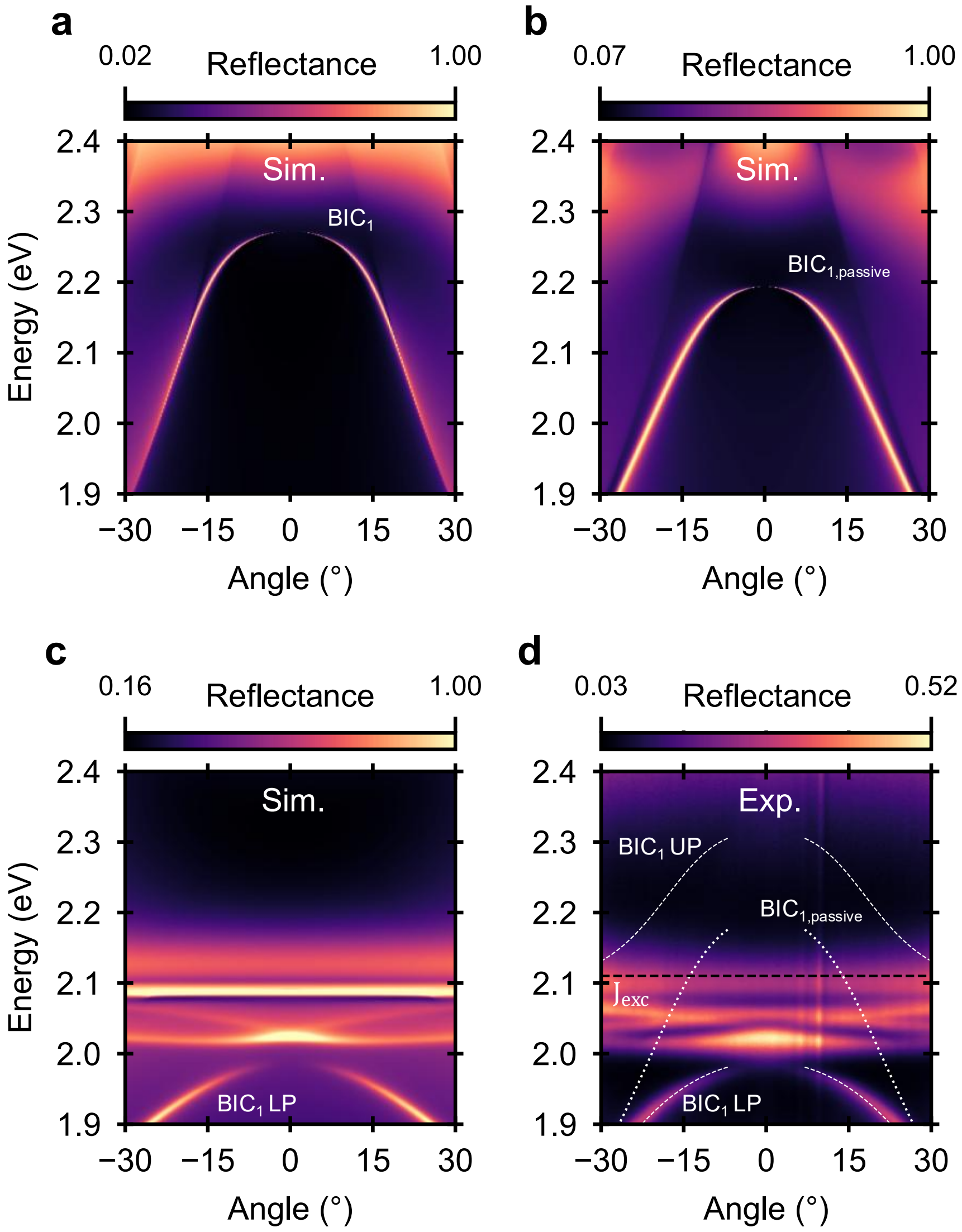

**Fig. S4: Calculated photonic dispersions of the metasurface BIC resonance. a-b,** Calculated angle-resolved reflectance of the bare (a), passive (**b**), and active (c) metasurface showing the typical dispersion of a BIC for the bare and passive metasurface ($BIC_1$ and $BIC_{1,passive}$, respectively), while the active metasurface shows significant redshift and distortion of the BIC band as expected from a lower polariton band ($BIC_1$ LP). **d**, Experimental angle-resolved reflectance of the active metasurface where the coupling between the J-exciton (black dashed line) and the passive BIC (white dotted line) results in the formation of the upper and lower polariton bands (white dashed lines). The upper and lower polaritons are calculated via the coupled oscillator model with a calculated Rabi splitting $\hbar\Omega_{LP\text{-}1}$ = 305.5±1.6 meV.

The Hopfield coefficients of the lower polaritons can be calculated as $|C_{LP}|^2 = \frac{1}{2}[1 - \frac{\Delta E}{\sqrt{\Delta E^2 + (\hbar\Omega)^2}}]$ and $|X_{LP}|^2 = \frac{1}{2}[1 + \frac{\Delta E}{\sqrt{\Delta E^2 + (\hbar\Omega)^2}}]$. The photonic fractions of the J-exciton-polaritons discussed in the manuscript are $|\mathrm{C}|^2{}_{\mathrm{BIC}_1 LP} = 0.40$, $|\mathrm{C}|^2{}_{\mathrm{BIC}_2 LP} = 0.47$, and $|\mathrm{C}|^2{}_{\mathrm{BIC}_3 LP} = 0.54$ (Fig. S5a).

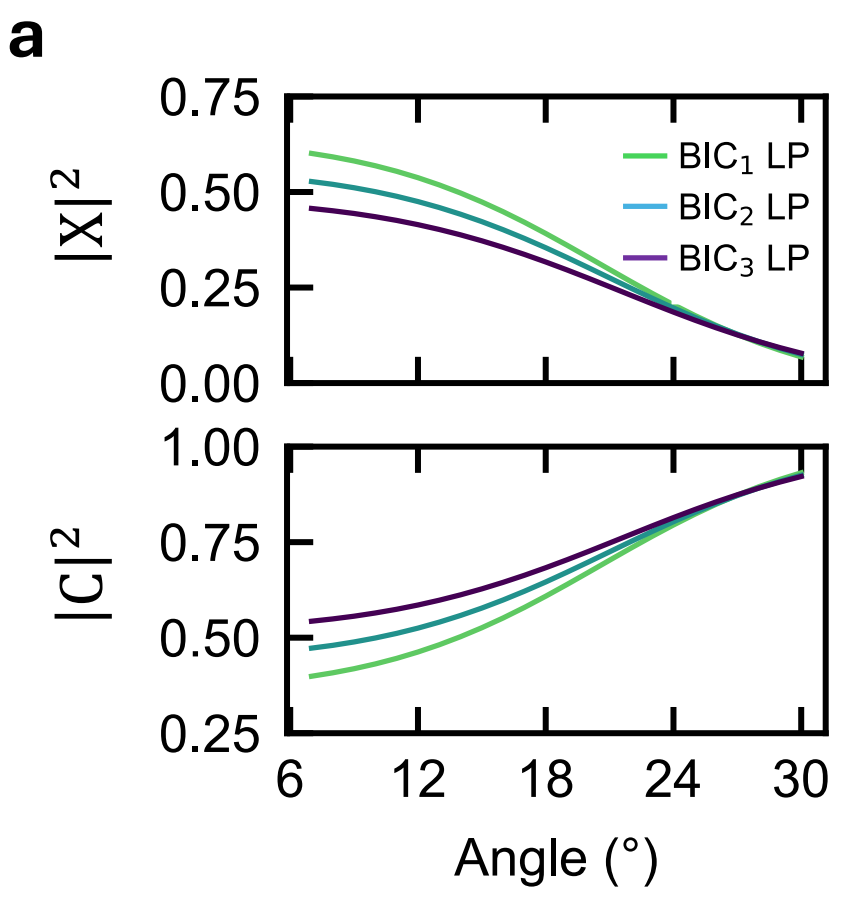

**Fig. S5: Lower polaritons Hopfield coefficients. a,** Excitonic ($|X|^2$) and photonic ($|C|^2$) fractions calculated for the three lower polaritons dispersion measured experimentally.

To understand where the J-aggregate film is positioned when deposited on the metasurface and how it affects strong coupling, we perform AFM measurements of a silicon metasurface covered by a thin film of J-aggregates and calculate the electric field intensity within the film when placed at different heights. An area scan of the covered metasurface (Fig. S6a) and the distribution of the heights in the same area (Fig. S6b) show that the J-aggregate film lays conformally to the silicon pillars and falls towards the substrate in-between pillars. Furthermore, taking a series of height profiles along the metasurface reveals that the J-aggregate retains uniformity of micrometer-scale distances (Fig. S6c).

The influence of the J-aggregate film position on coupling is quantified by the calculated electric field intensity (Fig. S6d). Thanks to the uniformity of the electric field distribution of the BIC along the z axis (Fig. S2c), the system enters the strong coupling regime and forms the BIC lower exciton-polariton around the same energy (2.0 eV in the geometry chosen for this example) irrespective of film position.

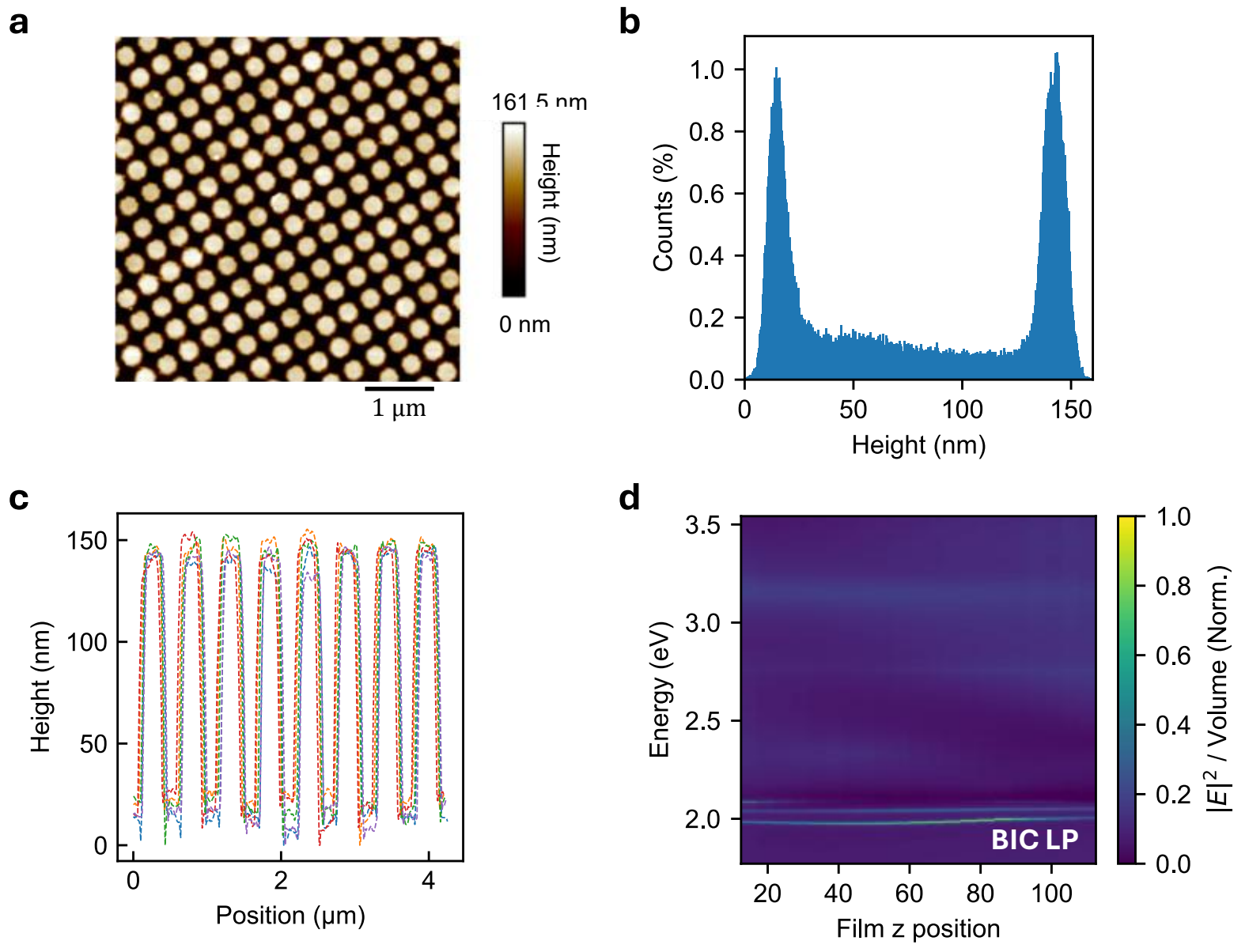

**Fig. S6: Nanoscale morphology of the J-exciton-polariton metasurface. a,** Atomic force microscopy area scan of a silicon metasurface covered by a J-aggregate thin film. **b,** Height distribution over the scanned area in **a**. **c,** Height profile obtained by extracting multiple linecuts from the scan area in **a**. **d,** Calculated electric field intensity in the J-aggregate film volume as a function of its vertical position along the silicon pillar. The BIC LP is shown to form at any of the vertical positions with minimal changes in its energy.

To evaluate whether in the strong coupled system local absorption enhancement effects are present at the excitation wavelength and whether they scale with Rabi splitting, we calculated the absorbed optical power density in the coupled J-aggregate film as $P_{abs} = \frac{1}{2}\varepsilon_0\omega\,|E|^2\,(2nk)$. For the three resonant metasurfaces in Fig. S7a, the absorbed optical power at the excitation energy (purple dashed line) is 59% higher than in the bare film but remains mostly unchanged between the three metasurfaces. This is consistent with the absence of an optical resonance at the excitation energy (Fig. S7b), where the electric field is only weakly confined within the TDBC film for all metasurfaces (Fig. S7c). Based on these results we conclude that local absorption enhancements induce negligeable effects on $g^{(2)}(0)$ and do not influence the superlinear growth of temporal correlations with excitation fluence.

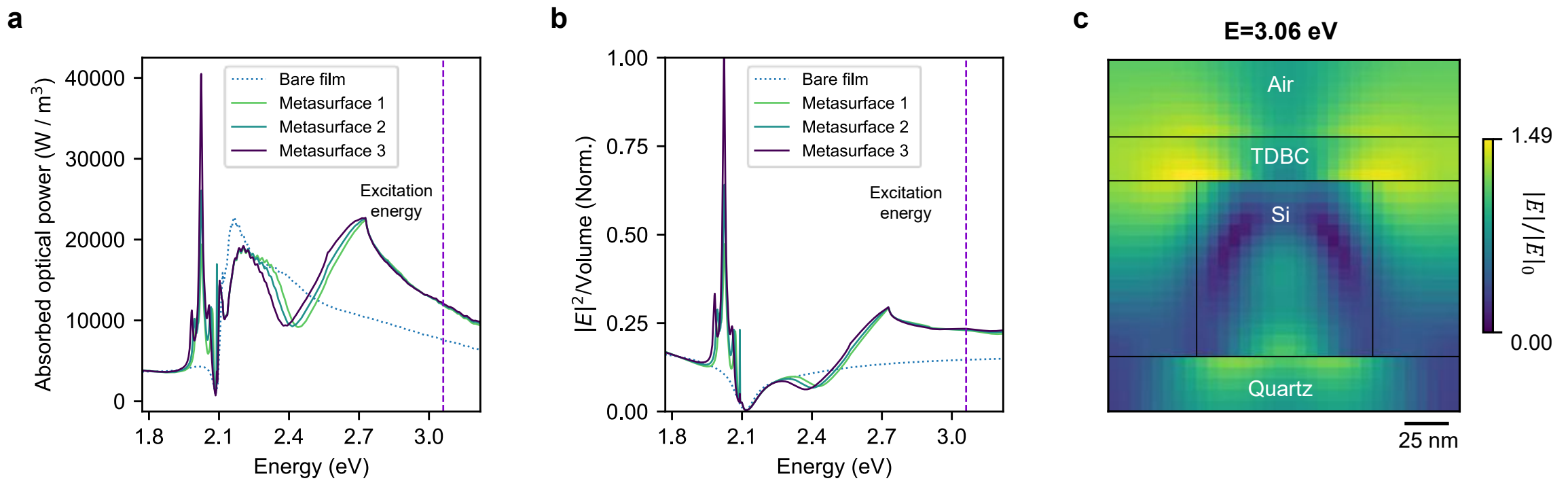

**Fig. S7: Simulated absorbed optical power of strongly coupled metasurfaces. a-b,** Rigorous coupled-wave analysis (RCWA) calculated absorbed optical power (**a**) and electric field (**b**) for the three metasurfaces coupled to the J-aggregate thin film showing no significant changes in absorption and electric field around the excitation energy (purple dashed line). **c,** Calculated electric field distribution in the yz plane at the excitation energy (3.06 eV) showing weak field confinement in the TDBC film.

## Supplementary Note 4: Superradiance in J-aggregates

Time-resolved photoluminescence measurements are performed to confirm the superradiant nature of J-excitons. The TDBC monomer powder was dispersed in methanol to prevent aggregation, and was then cast on glass via spin coating to form a film.

To characterize the dependence of superradiant emission on the number of excited J-aggregates, time-resolved measurements are performed as a function of excitation fluence. The peak intensity of the time-resolved PL of the bare J-aggregate film and of the lower polaritons is shown to grow super-linearly with excitation fluence (Fig. S8a), following a power law with exponents increasing from 1.12 to 1.55 along with the Rabi splitting and photonic fraction (dashed lines). The superlinear scaling of the emission peak intensity with pump power is a hallmark signature of superradiance[2–5], a direct consequence of the emission speed-up when an increasing number of emitters is participating in cooperative emission. While the increase in power law exponent with growing photonic fraction confirms the enhanced non-linearity induced by the metasurface. Furthermore, the lifetime of both the J-aggregate film and the lower polaritons (extracted by deconvolution fitting of the instrument response function calculated via the SymPhoTime 64 software) decreases as a function of excitation fluence (Fig. S8b), with the lower polaritons systematically having a shorter lifetime than the uncoupled J-aggregate emission. The emission speed-up observed in the system can thus be associated to two superradiant regimes[6]: *microscopic* superradiance, originating from the formation of delocalized excitons with a macrodipole of $N_{coh} = 6$ molecular units, and *macroscopic* superradiance arising from coherent coupling of distinct J-excitons.

a

b

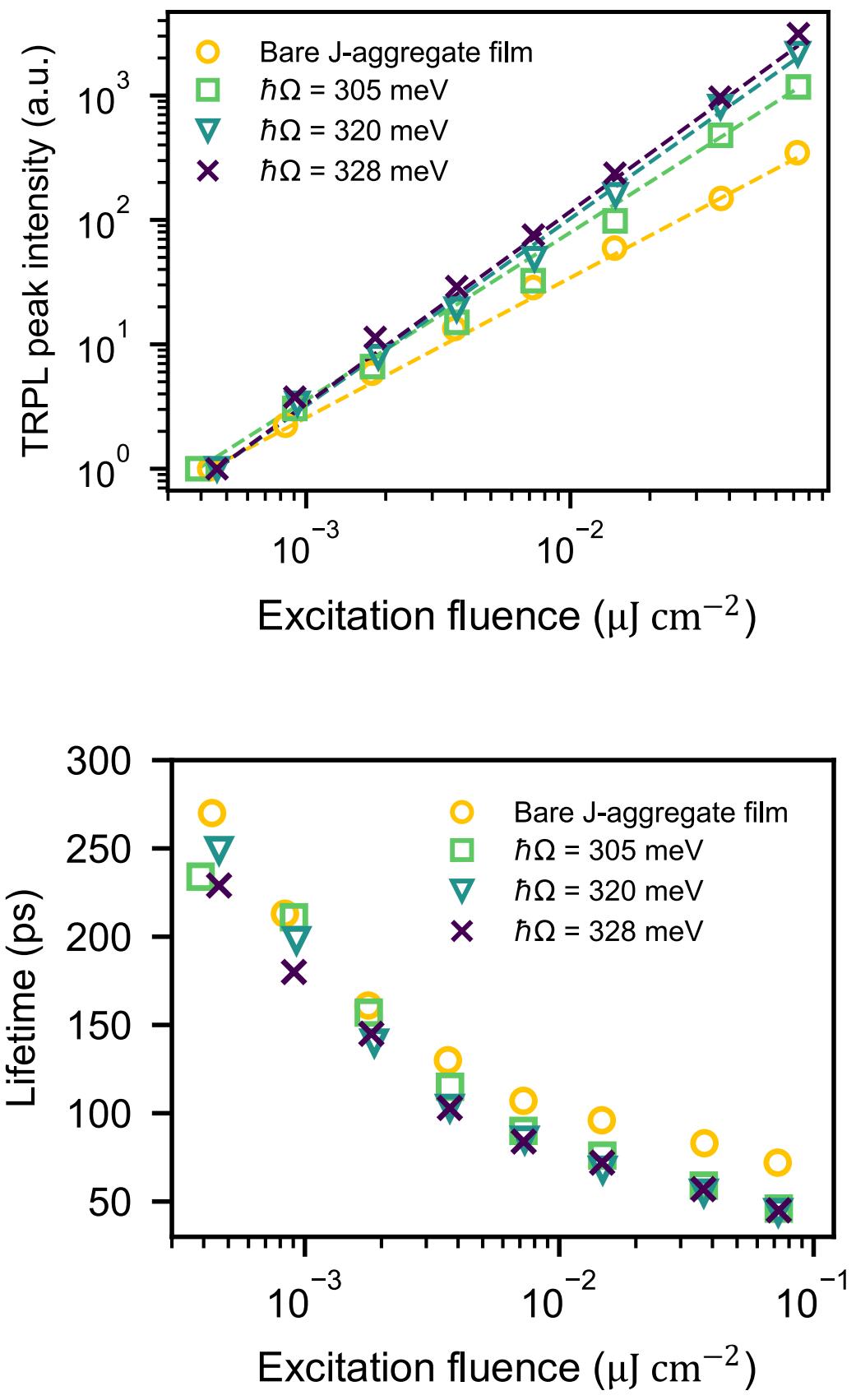

**Fig. S8: Superradiance in the time-resolved regime. a,** Normalized peak PL intensity in time-resolved measurements for the bare J-aggregate film and lower polaritons showing super-linear growth

with excitation fluence. The dashed lines represent a power law fit that yields n=1.12 and n=1.55 for the J-exciton and $BIC_3$ LP, respectively. **b,** Extracted lifetime of the J-exciton state (yellow circles) and lower polaritons demonstrating the typical power-dependent decay speed-up of superradiant emitters.

Considering that, for inter-aggregate superradiance, the radiative emission rate increases with excitation intensity, $k_r \propto I_{ex}$ , the steady-state emission intensity, $I_{PL}$, is

$$I_{PL} = \Phi I_{ex} = \frac{k_r}{k_r + k_{nr}} I_{ex} \propto \frac{1}{I_{ex} + k_{nr}} I_{ex}^2, \qquad \text{(S1)}$$

where Φ is the quantum yield and $k_r$ and $k_{nr}$ are the radiative and non-radiative decay rates, respectively. Reports of the scaling of steady-state emission intensity with excitation fluence for superradiant systems range from linear to quadratic[7–14], consistent with equation (S1) in the limit of $k_r(I_{ex}) \ll k_{nr}$ and in the presence of non-superradiant components with constant $k_r$. Considering the low quantum yield of TDBC ($\Phi_{\mathrm{TDBC}}$~0.02-0.05 [15]) and the presence of a long-lived component in the time-traces (Fig. 1c), the transition from linear to quadratic regime observed within the coupled systems indicates that the metasurface increases the superradiant component over the competing non-superradiant contributions.

Steady-state photoluminescence measurements provide similar evidence of enhanced cooperativity mediated by the strong light-matter interaction between J-excitons and the BIC mode (Fig. S9). In the coupled systems (Fig. S9b-d), the narrow luminescence peak associated to the LP branch redshifts relative to the bare J-aggregate film emission (Fig. S9a) following the Rabi splitting and coupling strength determined by the metasurface detuning. Linear to quadratic scaling of the luminescence with excitation intensity has been reported in various SR emitters[22,57–63]. This can be understood in terms of their intensity-dependent quantum yield in the different excitation regimes investigated, along with competing non-superradiant contributions. In our experiments, the steady-state PL intensity, $I_{PL}$, of the uncoupled J-aggregate film (Fig. S9e) increases linearly with excitation intensity, $I_{ex}$. In contrast, after accounting for metasurface-enhanced absorption at the excitation energy (see Fig. S7a), $I_{PL}$ of the lower polariton states exhibits a markedly nonlinear response with scaling $I_{PL} \propto I_{ex}^{n}$ (Fig. S9f-h), where the power law exponent $n$ grows from 1.7 to 2.0 as the Rabi splitting increases. This suggests that stronger exciton-cavity coupling enhances the superradiant response of the exciton-polaritons by increasing the population of superradiant states relative to that of all the other radiative states contributing to the emission.

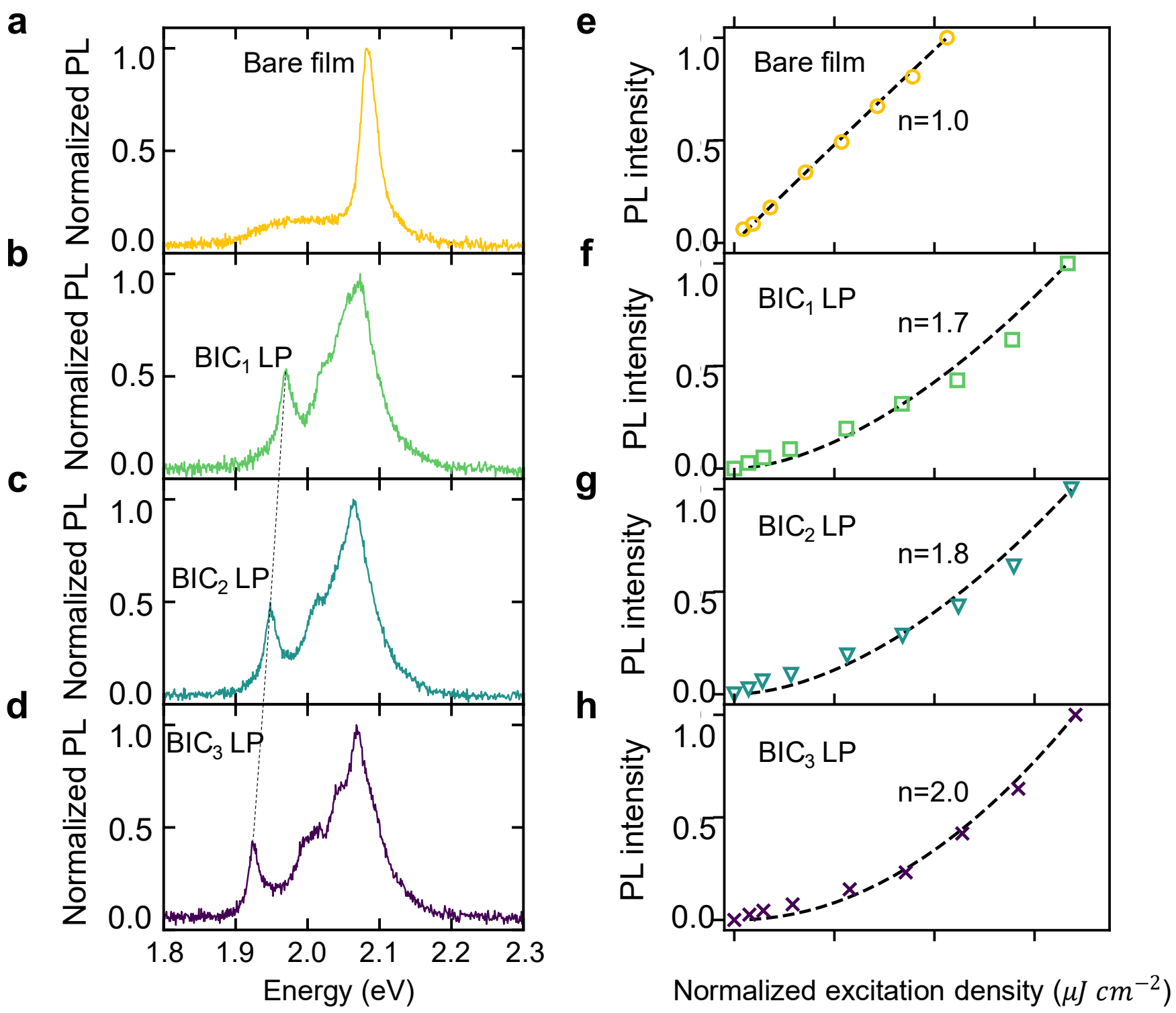

**Fig. S9: Superradiant photoluminescence spectra in the strong coupling regime. a-d**, Steady-state photoluminescence spectra of the J-aggregate film on quartz (**a**) and coupled to the three metasurfaces (**b-d**), showing a narrow LP peak that redshifts relative to the $J_{exc}$ emission as the Rabi splitting increases. **e–h**, Dependence of the integrated photoluminescence intensity of the J-exciton (**e**) and the lower polariton peak (**f-h**) on normalized excitation density. The J-exciton exhibits a linear dependence, while the lower polariton exhibits a distinct superlinear growth. The data are fitted with a power law (black dashed lines) with exponent $n$ that increases with Rabi splitting.

## Supplementary Note 5: Photon statistics of J-aggregates and Rhodamine-B films

Given the presence of multiple energy levels such as the J-excitons (2.11 eV), the self-trapped excitons (2.0 eV), and the BIC LPs (1.95-2.00 eV), $g^{(2)}(0)$ may be affected by auto- and cross-correlations between these states. Measurements of the J-exciton trap state emission show weakly varying, mostly uncorrelated, photon correlations as a function of fluence (Fig. S10a), suggesting that population funneling from the J-exciton to the trap state is not influencing photon statistics. Furthermore, the spectral overlap between the J-exciton and the BIC LPs emission does not increase systematically with coupling strength (Fig. S10b), excluding metasurface-enhanced interactions with self-trapped excitons as a mechanism for the increase in the observed $g^{(2)}(0)$. We note that all $g^{(2)}(\tau)$ measurements have been performed by filtering the emission of interest with a bandpass filter of FWHM~30 meV (Fig. S11).

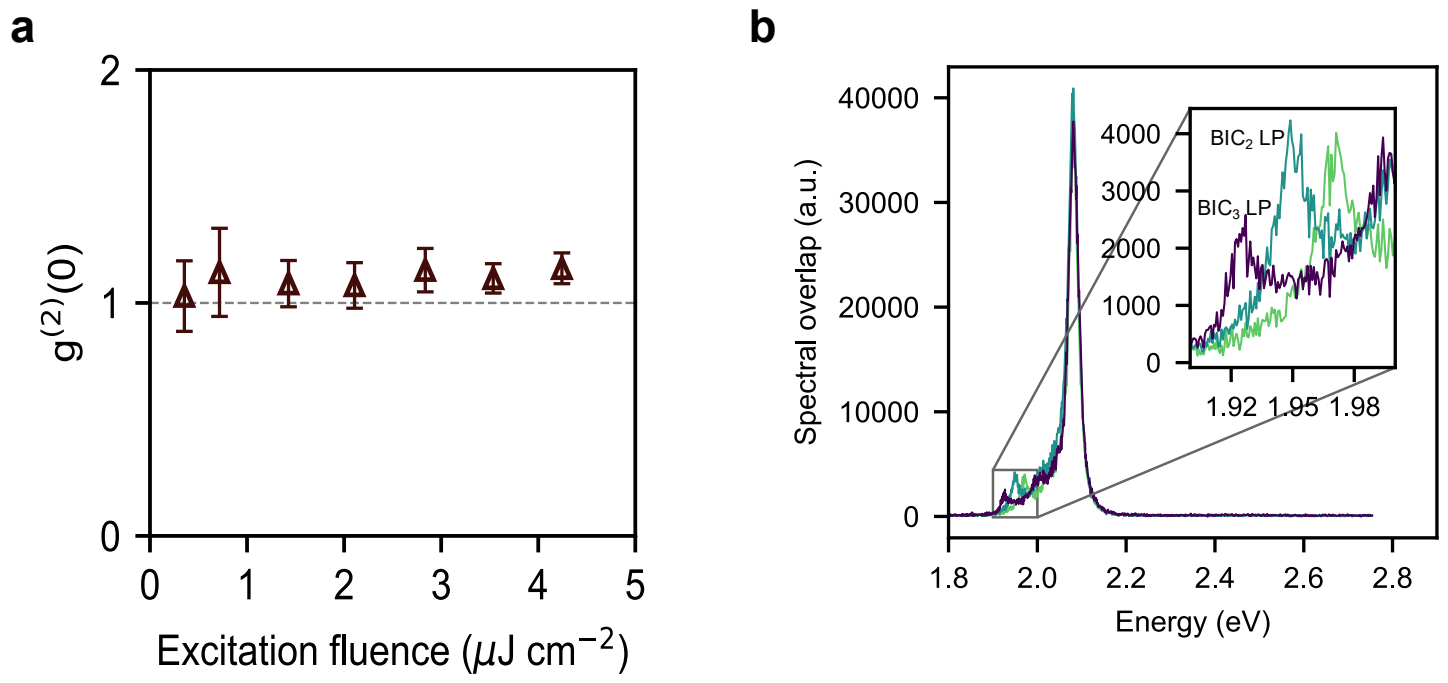

**Fig. S10: J-aggregate trap state temporal correlations. a,** Zero-delay second-order intensity correlation at increasing excitation fluences showing no significant increase in temporal correlation. **b,** Spectral overlap between the J-exciton and the BIC LPs emission showing no systematic increase with coupling strength.

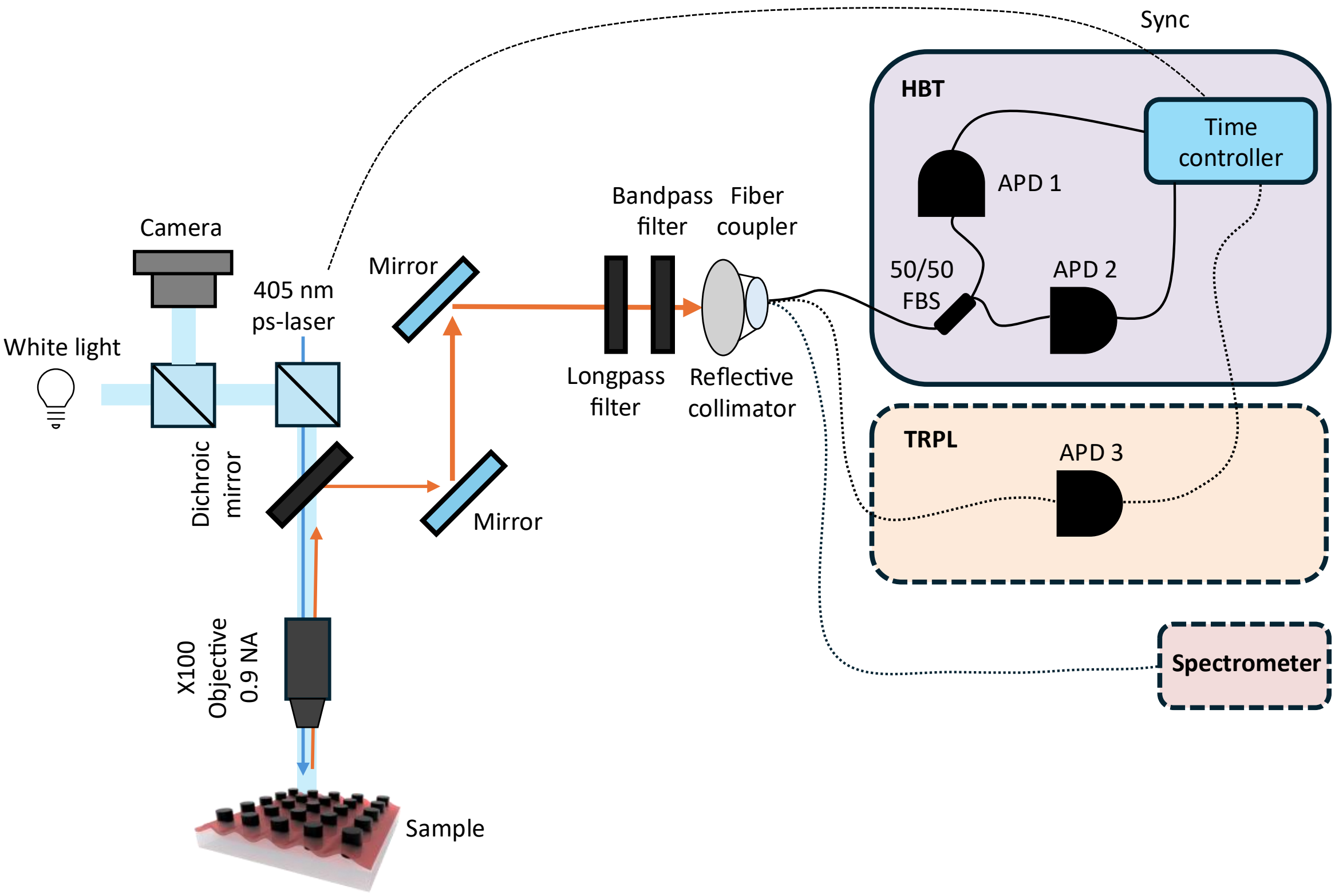

**Fig. S11: Photoluminescence characterization setup.** Schematic of the setup utilized for the characterization of the photoluminescence intensity, lifetime, and photon correlations. The collection path can be redirected via optical fibers to a spectrometer for PL spectral measurements, a time-controller for time-resolved measurements (TRPL), or to a Hanbury-Brown-Twiss setup (HBT) for photon correlation measurements.

A rhodamine-B film was also used as a control sample and its $g^{(2)}(\tau)$ was measured both for the bare film (Fig. S12a) and for a resonant metasurface covered by the film (Fig. S12b). In both cases, rhodamine-B does not display any significant changes in its photon statistics.

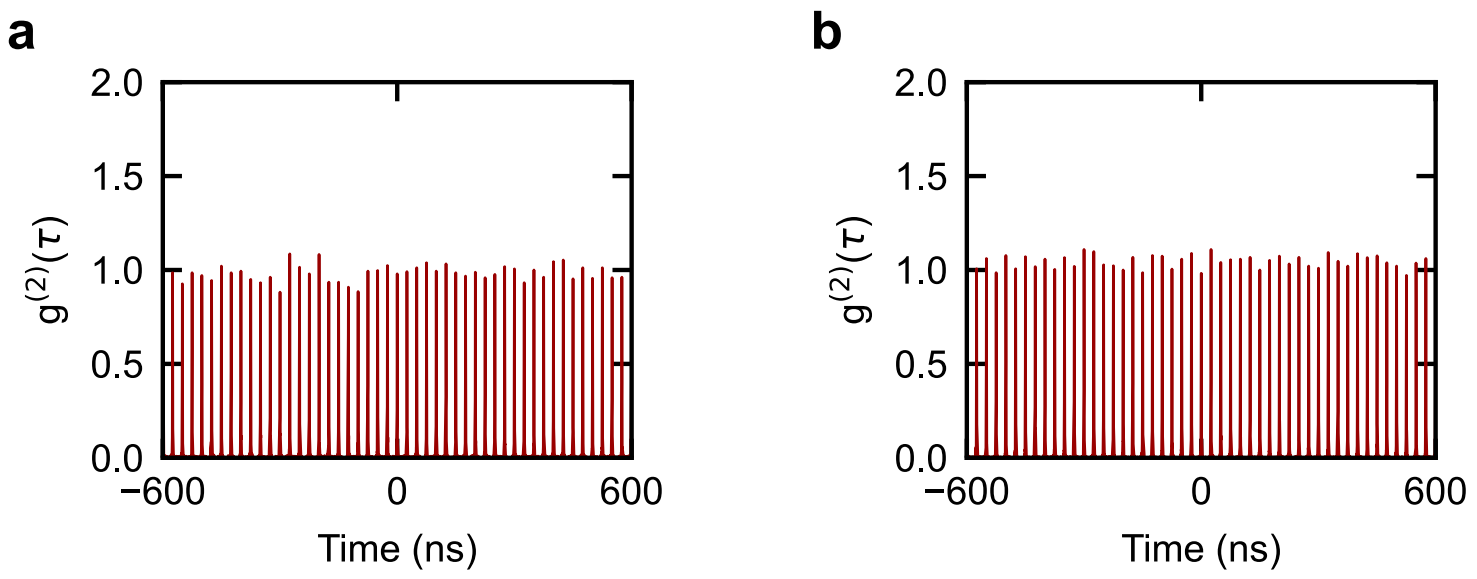

**Fig. S12: Intensity correlation measurements of Rhodamine-B film.** Intensity correlation function for a bare **a**, and a metasurface-coupled **b**, Rhodamine-B film, respectively. No change in photon correlations is observed after coupling to the metasurface.

## Supplementary Note 6: Modelling the second order correlation function in superradiant J-aggregates

Here we calculate the zero-delay second order correlation function defined as

$$g^{(2)}(0) = \frac{\mathrm{Tr}(\rho\hat{a}^\dagger\hat{a}^\dagger\hat{a}\hat{a})}{\left(\mathrm{Tr}(\{\rho\hat{a}^\dagger\hat{a}\})\right)^2},$$

where $\rho$ is the density matrix of the emitted light and $\hat{a}^\dagger$, $\hat{a}$ are the bosonic creation and annihilation operators.

Following the emission model proposed in the paper, we assume that the excitation volume is partitioned into *M* independent superradiant domains, each containing *N* single-photon emitters. Each domain emits a burst of *N* photons with probability *p*, and remains dark with probability *1-p*. Therefore, the total emission follows a binomial distribution with the density matrix

$$\rho = \sum_{m=0}^{M} \binom{M}{m} p^m (1-p)^{M-m} |Nm\rangle\langle Nm|,$$

where the term with index *m* corresponds to emission of *Nm* photons from *m* domains. This allows us to calculate the average number of photons as

$$\mathrm{Tr}(\rho\hat{a}^\dagger\hat{a}) = \sum_{m=0}^{M} Nm \binom{M}{m} p^m (1-p)^{M-m}.$$

Using the identity $m\binom{M}{m} = M\binom{M-1}{m-1}$ and substituting *m-1=k*, we obtain

$$\mathrm{Tr}(\rho\hat{a}^\dagger\hat{a}) = pNM \sum_{k=0}^{M-1} \binom{M-1}{k} p^k (1-p)^{M-1-k} = pNM.$$

Similarly, we get the numerator of the correlation function as

$$\mathrm{Tr}(\rho\hat{a}^{\dagger}\hat{a}^{\dagger}\hat{a}\hat{a}) = \sum_{m=0}^{M} Nm(Nm-1)\binom{M}{m}p^{m}(1-p)^{M-m} =$$

$$= N^2p^2M(M-1) + pNM(N-1).$$

Substituting these results into the definition of $g^{(2)}(0)$ gives

$$g^{(2)}(0) = \frac{N^2p^2M(M-1)+pNM(N-1)}{(pNM)^2} = \frac{M-1}{M} + \frac{N-1}{pNM}.$$

This compact expression shows how temporal photon bunching arises from the interplay of the number of domains *M*, the number of emitters per domain *N*, and the burst probability *p*.

It should be noted that this expression describes normalized photon-number fluctuations rather than the absolute strength of radiative emission. Accordingly, even for a single superradiant domain, $g^{(2)}(0)$ does not necessarily increase with increasing radiative efficiency. In the limit $\Phi \rightarrow 1$, the emission becomes more deterministic from shot to shot, reducing the relative intensity fluctuations and therefore suppressing the normalized bunching. Conversely, in the limit of small Φ, emission occurs through rarer but more intermittent cooperative events, so the variance remains large relative to the mean and $g^{(2)}(0)$ correspondingly increases. The increase of $g^{(2)}(0)$ in this regime therefore reflects enhanced fluctuation statistics, not a larger total photon yield.

**Supplementary Note 7: Large scale J-exciton-polariton spatial coherence**

Spatial coherence measurements of the bare J-aggregate film and the J-exciton-polaritons emission (Fig. 5d-f) are measured via a Michelson interferometer (see Methods for more details on the experimental setup). At the highest excitation density ($\mathrm{P} = 7.03\ \mu\mathrm{J\ cm^{-2}}$) the J-aggregate film does not show any evidence of interference fringes formation (Fig. S13a). In contrast, the emission of all BIC LPs show distinct vertical fringes in its real space image (Fig. S13b-d). Taking the Fourier transform of the real space emission shows that no high frequency side lobes are present for the J-aggregate film, while horizontal side lobes are visible for all BIC LPs.

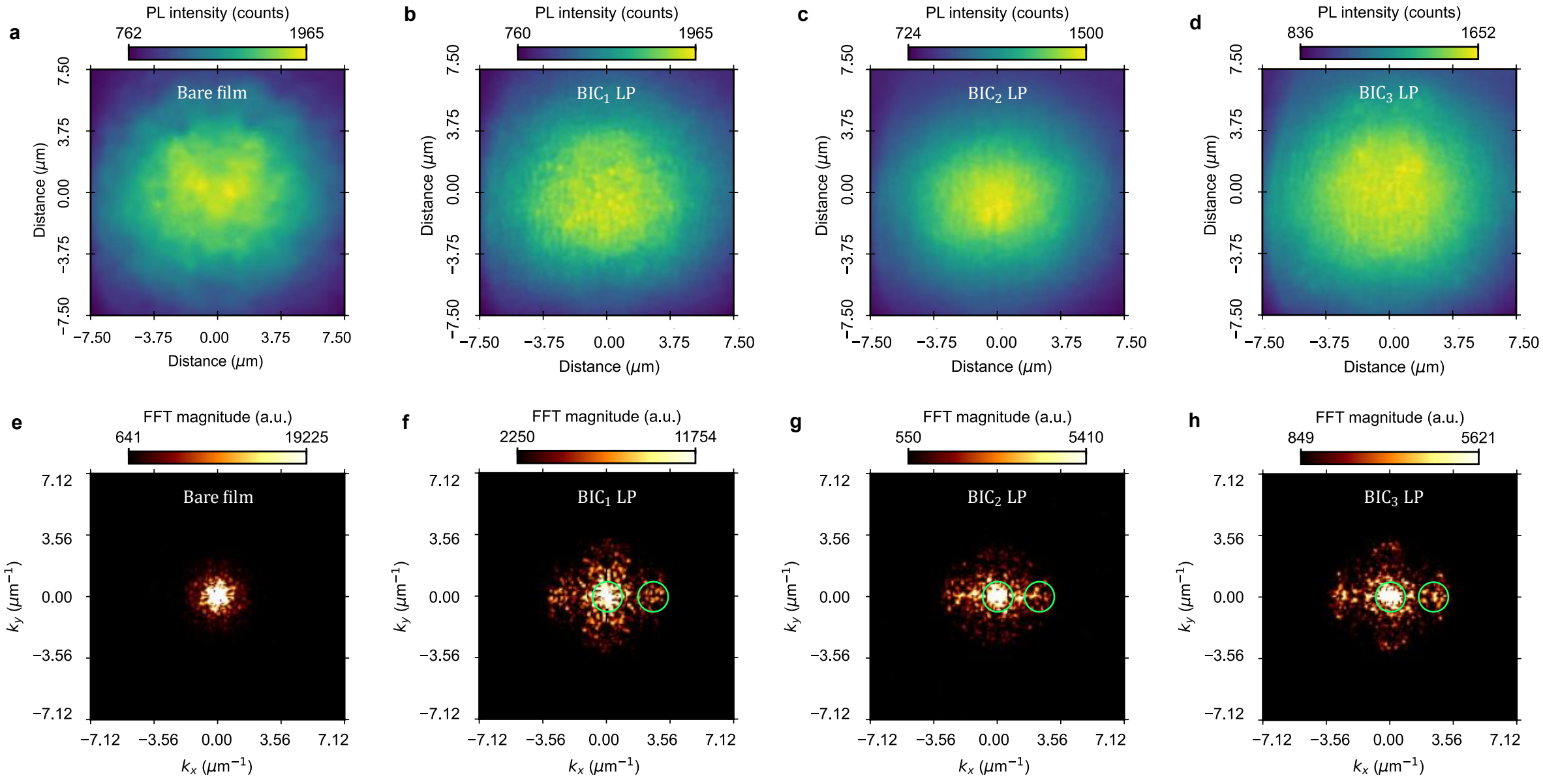

**Fig. S13: Interference fringe formation in strongly coupled J-aggregates. a,** Real space image of the uncoupled J-aggregate emission and **b-d,** the lower polariton emission, showing no measurable spatial coherence for the uncoupled J-aggregate and fringe formation for the lower polariton emission. **e-h**, The corresponding Fourier plane show no side lobes for the bare J-aggregate emission, while horizontal high frequency components corresponding to the interference fringes in **b-d** are present for all BIC LPs (green circles).

To quantitatively study the evolution of the superradiant domains with fluence and photonic fraction we used a combined amplitude- and phase-based domain selection algorithm, assuming that physically meaningful phase domains should occur only in regions where the complex field amplitude is sufficiently large and the phase is sufficiently flat. The amplitude maps are used to define the spatial region where the phase signal is reliable, while the wrapped phase maps are used to determine whether each candidate region is internally phase coherent.

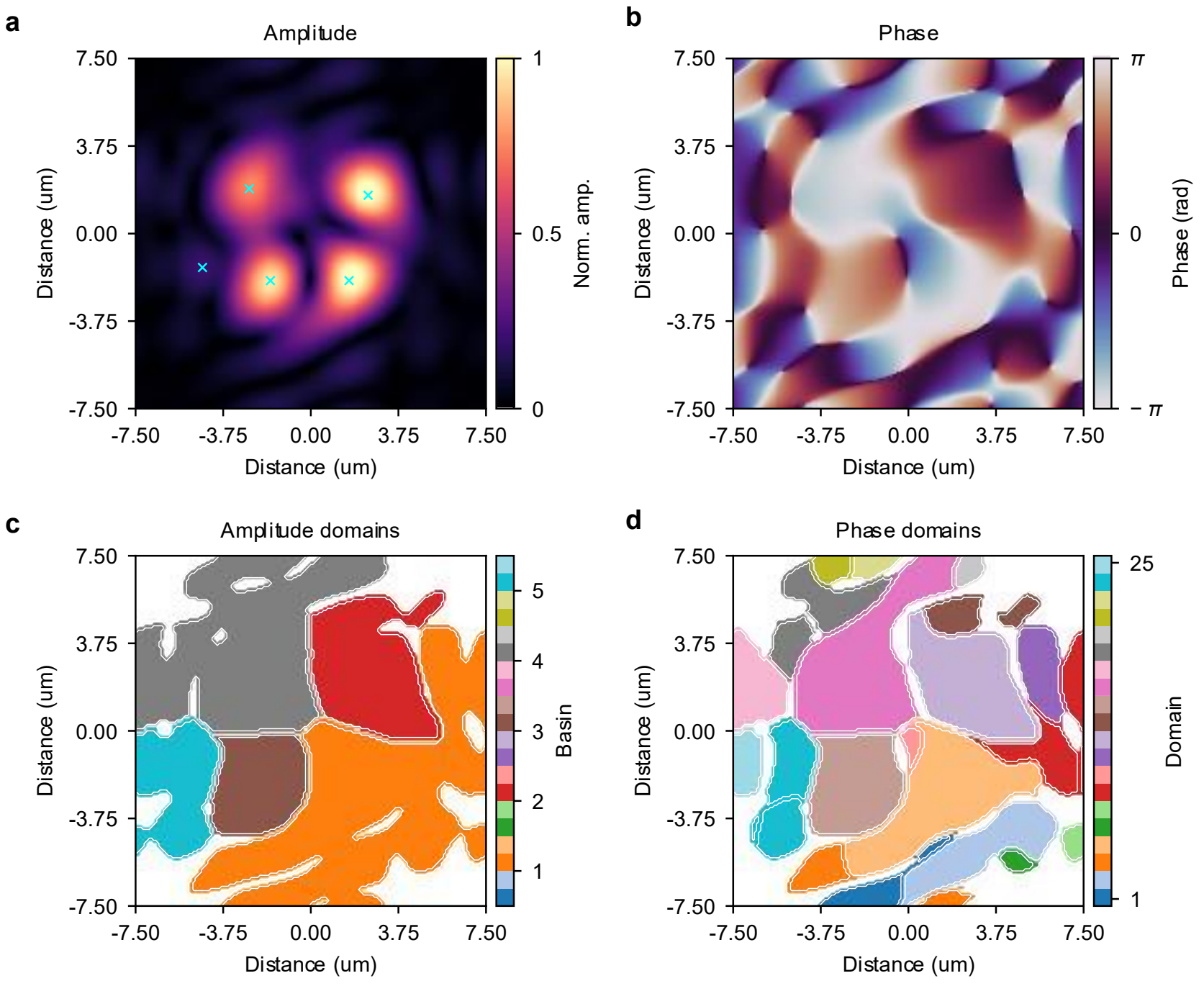

**Fig. S14: Domain classification algorithm. a**, Amplitude and **b,** phase maps of one of the interference patterns analysed in the manuscript showing both quantities varying within the emitting region. **c,** Amplitude selection step dividing different spatial region into amplitude basins. **b,** Phase classification step further dividing the amplitude basins into areas of slow varying phase separated by large phase shifts.

The algorithm runs as follows:

1. Pixels with low or invalid amplitude are rejected using an amplitude threshold (20th percentile), so that phase values in noise-dominated regions are not interpreted as physical domains;
2. The remaining regions are scanned for local maxima, and each maximum is used as a seed to define amplitude basins;
3. Within each amplitude basin, pixels are grouped into phase-compatible areas (each pixel must differ $< \frac{\pi}{4}$ from the previous pixel);
4. Candidate domains are then filtered by minimum area (set as 1 $\mu m^2$, approximately the minimum size of a fringe period observed experimentally) and a phase-constancy criterion requiring the phase distribution within the domain to remain sufficiently uniform (the phase deviation must be $< \frac{\pi}{4}$ from the mean phase of the domain);

5. Accepted spatial domains are grouped into phase families, allowing disconnected same-phase fragments to be distinguished from out-of-phase domains (regions with $> \frac{\pi}{4}$ phase shifts).

An example of the domain classification algorithm is shown in Fig. S14, where prior to processing the amplitude (Fig. S14a) and phase (Fig. S14b) maps show a textured distribution of both amplitude and phase islands. In the first instance, the amplitude selection step identifies 5 large amplitude basins separated by significant drops in amplitude (Fig. S14c), while the phase selection step divides those basins into smaller domains defined by internally slowly varying phase and surrounded by large phase shifts (Fig. S14d). The areas where the amplitude is significant with respect to the background and where the phase is sufficiently constant are classified as valid domains for further size analysis.

Performing this analysis for the amplitude and phase maps of $BIC_3LP$ allows extracting the number of emissive domains (Fig. S15a) and the domain area survival distribution (Fig. S15b) discussed in detail in the manuscript.

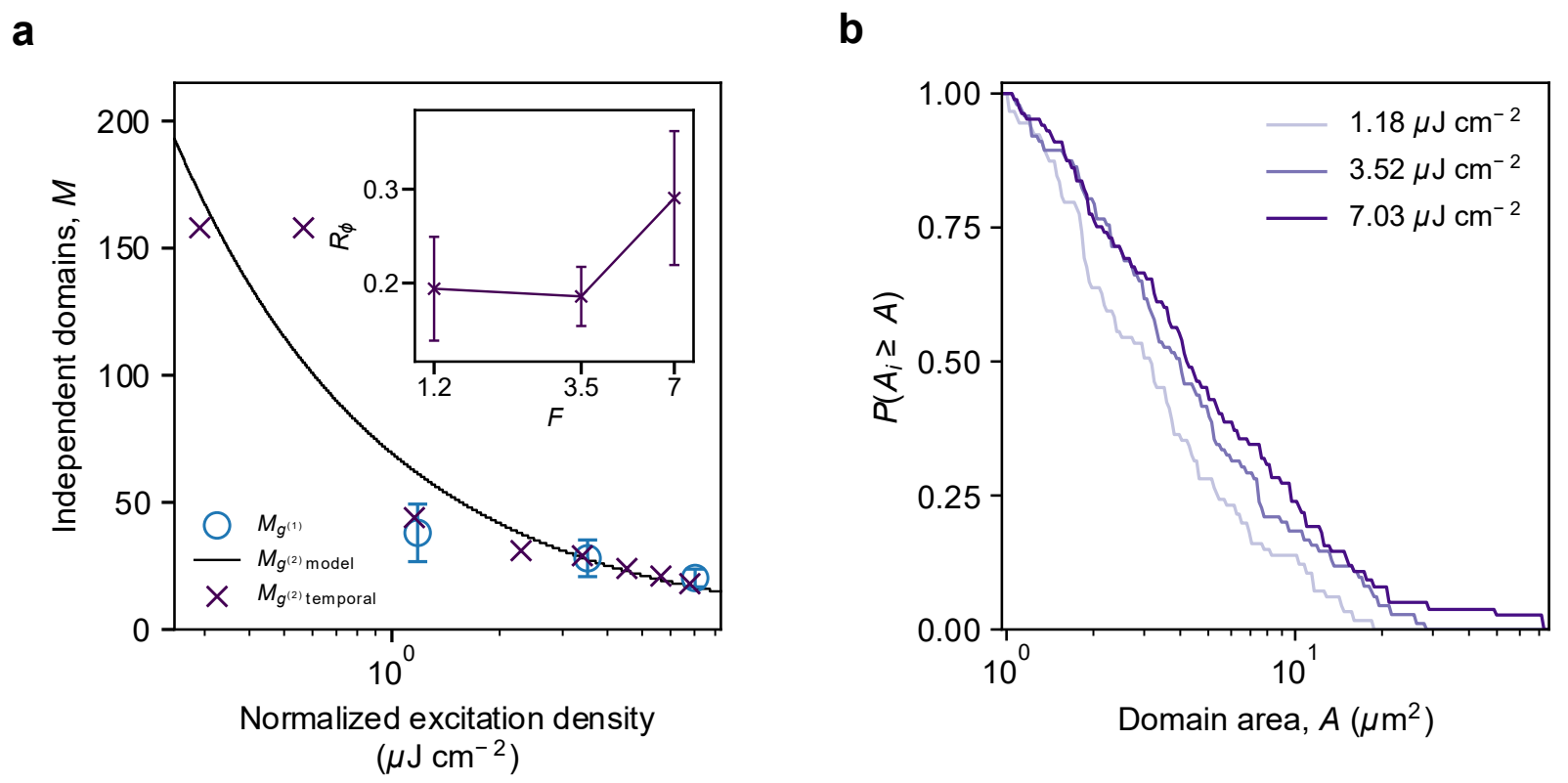

**Fig. S15 Metasurface-enhanced superradiant clustering. a,** Number of independent domains, $M$, extracted from the BIC $LP_3$ amplitude and phase maps (blue circles) compared with the values from the $g^{(2)}$ model (black continuous line and purple crosses) as a function of excitation density. Inset: global phase-order parameter, $R_{\phi}$. Error bars indicate variability across different metasurface locations. **b,** Domain-area survival probability at three excitation densities.